\documentclass[aip,graphicx]{revtex4-2}

\usepackage{nicefrac} 
\usepackage{xcolor}
\usepackage{amsmath, amsthm, amssymb} 
\usepackage{graphicx}
\usepackage{dcolumn}
\usepackage{bm}
\usepackage{hyperref} 

\usepackage[utf8]{inputenc}
\usepackage[T1]{fontenc}
\usepackage{mathptmx}
\usepackage{etoolbox}
\usepackage{placeins} 
\usepackage{subfig}
\usepackage{caption}
\usepackage{comment}
\usepackage{float}
\definecolor{MyGreen}{rgb}{0.45,0.60,0.09}

\usepackage{xr}

\begin{document}


\title{Asymptotically-consistent analytical solutions for the non-Newtonian Sakiadis boundary layer}  



\author{Nastaran Naghshineh}
\email[corresponding author: ]{nxncad@rit.edu}
\affiliation{School of Mathematical Sciences, Rochester Institute of Technology, Rochester, NY, 14623, USA}
\affiliation{Department of Sciences and Liberal Arts, Rochester Institute of Technology-Dubai, Dubai, 341055, UAE}

\author{Nathaniel S. Barlow}
\affiliation{School of Mathematical Sciences, Rochester Institute of Technology, Rochester, NY, 14623, USA}

\author{Mohamed A. Samaha}
\affiliation{School of Mathematical Sciences, Rochester Institute of Technology, Rochester, NY, 14623, USA}
\affiliation{Department of Mechanical and Industrial Engineering, Rochester Institute of Technology-Dubai, Dubai, 341055, UAE}
        
\author{Steven J. Weinstein}
\affiliation{School of Mathematical Sciences, Rochester Institute of Technology, Rochester, NY, 14623, USA}

\affiliation{Department of Chemical Engineering, Rochester Institute of Technology, 
Rochester, NY, 14623, USA}



\date{\today}
\begin{abstract}
{The Sakiadis boundary layer induced by a moving wall in a semi-infinite fluid domain is a fundamental laminar flow field relevant to high speed coating processes. This work provides an analytical solution to the boundary layer problem for Ostwald-de Waele power law fluids via a power series expansion, and extends the approach taken for Newtonian fluids ["On the use of asymptotically motivated gauge functions to obtain convergent series solutions to nonlinear ODEs", IMA J. of Appl. Math., (2023)] in which variable substitutions (which naturally determine the gauge function in the power series) are chosen to be consistent with the large distance behavior away from the wall. Contrary to prior literature, the asymptotic behavior dictates that a solution only exists in the range of power law exponents, $\alpha$, lying in the range $0.5 < \alpha  \leq 1$. An analytical solution is obtained in the range of approximately $0.74 \leq \alpha < 1$, using a convergent power series with an asymptotically motivated gauge function. For power laws corresponding to $0.5 < \alpha<0.74$, the gauge function becomes ill-defined over the full domain, and an approximate analytical solution is obtained using the method of asymptotic approximants ["On the summation of divergent, truncated, and underspecified power series via asymptotic approximants", Q. J. Mech. Appl. Math., (2017)]. The approximant requires knowledge of two physical constants, which we compute a priori using a numerical shooting method on a finite domain. The utility of the power series solution is that it can be solved on the entire semi-infinite domain and--in contrast to a numerical solution--does not require a finite domain length approximation and subsequent domain length refinement.}
\end{abstract}
\maketitle 
\section{Introduction \label{sec:Intro_NNS}}
The Sakiadis boundary layer\cite{Sakiadis} is a fundamental flow field in processes where laminar liquid films are coated onto moving substrates~\cite{Weinstein2004}. One of its key physical implications is in the area of high speed curtain coating, where the boundary layer length is essential to the  mechanism of hydrodynamic assist that can suppress air entrainment~\cite{Blake}. In particular, its length determines where the wetting line is located with respect to the main body of the curtain flow.  Depending on the relative speed of the substrate and curtain flow at its bottom, the wetting line can lie directly underneath the curtain or can be dragged forward (lower curtain flow and higher substrate speeds) or retarded backward (high curtain flow and lower substrate speed). This wetting line location determines whether the stagnation pressure from a tall liquid curtain is sufficient to suppress the creation of an unstable air-bearing that leads to uneven and bubble-laden coatings. The highest coating speeds occur when the wetting line is located directly underneath the centerline of the curtain itself. The taller the liquid curtain, the faster the ultimate coating speed, provided that the wetting line location--again dictated by the Sakiadis boundary layer--is optimally controlled~\cite{Blake}. In addition to its relevance to coating, the Sakiadis boundary layer forms the basis for many studies including moving elastic sheets involving various modes of heat and mass transport\cite{Paper1, Paper2, Paper3, Paper4,Paper5, Paper6, Paper7}; in the past 5 years, the original Sakiadis paper~\cite{Sakiadis} has been cited over 400 times which demonstrates its continued fundamental importance. 

\begin{figure}[htbp]
  \centering
  \includegraphics[width=14cm]{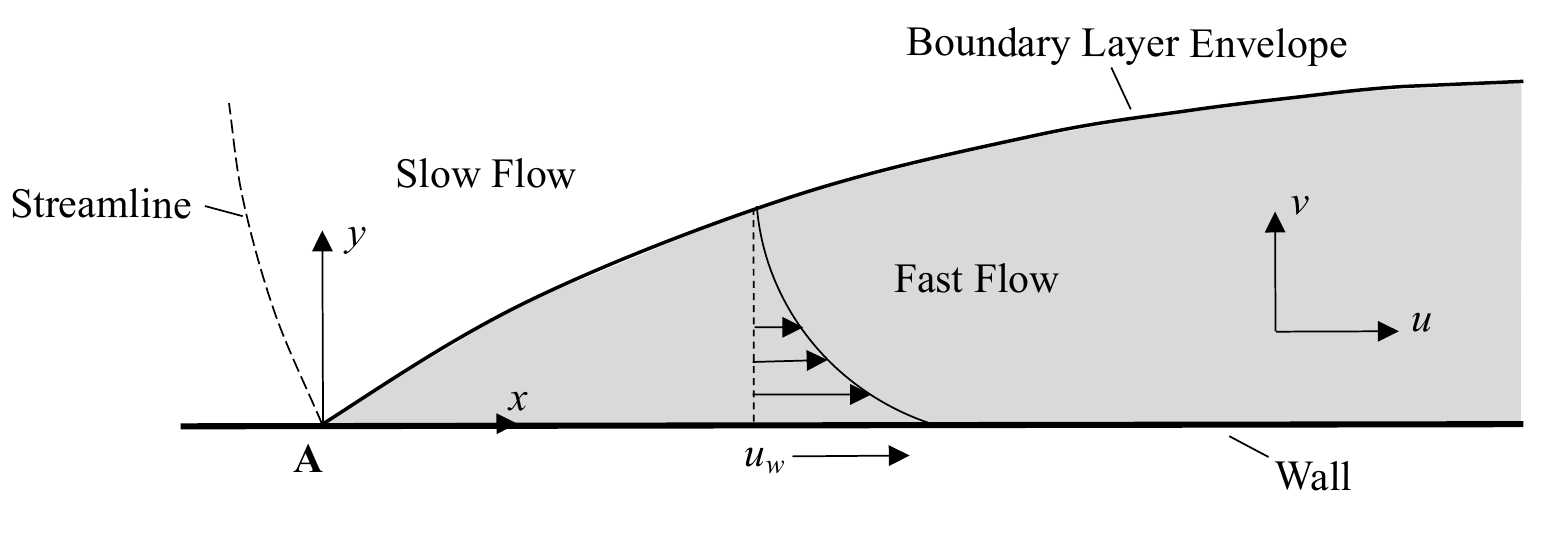}
    \caption{Schematic of the Sakiadis boundary layer flow. Slowly moving or otherwise stationary fluid with characteristic velocity scale $S$ is in contact with a fast moving wall having speed of $u_w$ as indicated, where $S<<u_w$. As a result, the velocity $u \to 0 $ as $y \to \infty$ in the boundary layer approximation to the flow equations. The boundary layer envelope is defined in this paper as the locus of points for which $u/u_w=0.1$. In the original papers of Sakiadis~\cite{Sakiadis} and Fox et al.~\cite{Fox}, fluid enters the domain at point $\bold{A}$ through a slit, where the streamline is redrawn to be vertical and coincident with a wall. In coating applications, the streamline often aligns with an interface where point $\bold{A}$ is a moving contact line~\cite{Weinstein2004}.} 
  \label{fig:Sakiadis_Config}
\end{figure}

Figure~\ref{fig:Sakiadis_Config} shows the configuration of the Sakiadis boundary layer problem with the $x$-$y$ coordinate system as indicated; the fluid flow is assumed to be invariant with the direction oriented out of the figure. Here, a flat wall is moving with velocity, $u_w$, through an otherwise stationary generalized Newtonian incompressible fluid of density, $\rho$, and viscosity, $\mu$.

The governing equations embody conservation of mass and momentum through the two dimensional incompressible steady continuity and Navier-Stokes equations. At high substrate speeds, velocity in the $x$-direction, $u$, is much larger than that in the $y$-direction, $v$, and velocity gradients in the $y$-direction dominate viscous forces in the boundary layer. These assumptions, which result in a small slope of fluid points, lead directly to Prandtl’s boundary layer equations—a nonlinear partial differential equation (PDE) system—that apply to the Sakiadis flow. For fluids having a viscosity that satisfies the Ostwald-de Waele power law dependence, the Sakiadis boundary layer equations are expressed as~\cite{Fox}
\begin{subequations}
\begin{equation}
    \frac{\partial u}{\partial x} + \frac{\partial v}{\partial y} = 0,
    \label{eq:ContinuityEq_NNS}
\end{equation}
\begin{equation}
    u \frac{\partial u}{\partial x}  + v \frac{\partial u}{\partial y} = \frac{1}{\rho}\frac{\partial \tau_{xy}}{\partial y},
\end{equation}
\begin{equation}
    \tau_{xy} = \mu \left(\frac{\partial u}{\partial y}\right)~,~\mu = K\left(-\frac{\partial u}{\partial y}\right)^{\alpha-1},
\end{equation}
\begin{equation}
    u = u_w,~v = 0 ~\textrm{at} \ y = 0~;~u \to 0~\textrm{as} \  y \to \infty.
\end{equation}
\label{eq:FoxEquations_NNS}
\end{subequations}
In~(\ref{eq:FoxEquations_NNS}c), $\tau_{xy}$ is the shear stress in the fluid, $K>0$ is the consistency coefficient, $\alpha$ is the power law exponent, and $\mu$ is the strain-rate dependent viscosity. Note that the rate of strain in the boundary layer approximation, $\partial u/\partial y$, is negative in the flow domain; thus, the magnitude of the rate of strain invokes a negative sign as indicated in the viscosity expression in~(\ref{eq:FoxEquations_NNS}c). Through the use of the stream function, $\psi$, that satisfies the continuity equation~(\ref{eq:ContinuityEq_NNS}) (i.e., $u = \partial \psi/ \partial y$ and $v = -\partial \psi / \partial x$), Fox et al.~\cite{Fox} define similarity variables given as 
\begin{subequations}
\begin{equation}
    \eta = y \left(\frac{\rho u_w^{2-\alpha}}{Kx}\right)^{1/(\alpha+1)},
    \label{eq:eta}
\end{equation}
\begin{equation}
    \psi = \left(\frac{y~u_w}{\eta}\right)f(\eta),
    \label{eq:psi}
\end{equation}
\label{eq:psi&eta}
\end{subequations}
and thus $u$ and $v$ are expressed as
\begin{equation}
    u=u_w\frac{df}{d\eta}~,~~v=\frac{1}{1+\alpha}\left(\frac{Ku_w^{2\alpha-1}}{\rho x\alpha}\right)^{\frac{1}{\alpha+1}}\left(\eta\frac{df}{d\eta}-f\right).
\label{eq:u}
\end{equation}
Upon substitution of~(\ref{eq:psi&eta}) into the system~(\ref{eq:FoxEquations_NNS}) and after rearrangement, Fox et al.~\cite{Fox} obtain the non-Newtonian Sakiadis boundary layer problem given as
\begin{subequations}
\begin{equation}
    \alpha(\alpha+1)\frac{d^3f}{d\eta^3} - (-\frac{d^2f}{d\eta^2})^{(2-\alpha)}f=0,~~~0\le\eta<\infty,
    \label{eq:ODE_NNS}
\end{equation}
\begin{equation}
    f = 0~\textrm{at} \ \eta=0,
    \label{eq:f(0)BC_SP_NNS}
\end{equation}
\begin{equation}
    \frac{df}{d\eta}= 1~\textrm{at} \ \eta=0,
    \label{eq:f'(0)BC_SP_NNS}
\end{equation}
\begin{equation}
     \frac{df}{d\eta}= 0~\textrm{as} \ \eta \to \infty.
     \label{eq:f'(inf)BC_SP_NNS}
\end{equation}
\label{eq:ODESystem_NNS}
\end{subequations}
Note that for a Newtonian fluid ($\alpha=1$), the equation system~(\ref{eq:ODESystem_NNS}) is identical to that of Blasius~\cite{Blasius} except that the location of conditions~(\ref{eq:f'(0)BC_SP_NNS}) and~(\ref{eq:f'(inf)BC_SP_NNS}) are reversed\cite{Sakiadis}. Nevertheless, the solution of~(\ref{eq:ODESystem_NNS}) is not simply related to that of Blasuis by translation, owing to its nonlinear governing equation~(\ref{eq:ODE_NNS}). 
 
Analytical solutions to the Newtonian Sakiadis problem have been examined by Barlow et al.~\cite{Barlow:2017}, who show that a power series solution about $\eta = 0$ has a finite radius of convergence and cannot bridge the physical domain $\eta \in [0, \infty)$. Naghshineh et al.~\cite{FlatWallSakiadisPaper} provide a convergent power series solution for the Newtonian problem in terms of exponential gauge functions consistent with the asymptotic behavior of the solution away from the wall; this behavior is given as
\begin{equation}
    f(\eta) \sim C + \tilde{a}_1e^{-C\eta/2}+ \tilde{a}_2 \left(e^{-C\eta/2}\right)^2 + O\left( \left(e^{-C\eta/2}\right)^3\right)~\textrm{as} \  \eta \to \infty,
    \label{eq:AsymptoticBehav_NS}
\end{equation}
where $C>0$ (this is consistent with the wall motion inducing a net volumetric flow in the positive $x$-direction in Fig.~\ref{fig:Sakiadis_Config}), $\tilde{a}_1$ and $\tilde{a}_2$ are asymptotic constants, and by inspection,
\begin{equation}
    \lim_{\eta \rightarrow \infty}f(\eta) \equiv C.
    \label{eq:AsymsptoticC_NNS}
\end{equation}
A Taylor series solution is given as
\begin{subequations}
\begin{equation}
    g(\omega) = \sum_{n=0}^\infty \tilde{a}_n \omega^n,
    \label{eq:omega}
\end{equation} 
\begin{equation}
     \omega(\eta) = e^{\displaystyle -C\eta/2},
     \label{eq:SakTransDefs}
\end{equation}
\begin{equation}
    f(\eta) \equiv  g(\omega(\eta)),
    \label{eq:composition}
\end{equation}
\label{eq:Solution_NS}
\end{subequations}
where all coefficients $\tilde{a}_n$ are provided by Naghshineh et al.~\cite{FlatWallSakiadisPaper} As written, the solution~(\ref{eq:Solution_NS}) uses $\eta = \infty$ as an expansion point; a slightly faster converging Taylor series expansion (in this transformed gauge function\footnote{A gauge function is the usual independent variable in a series expansion.  For example, in $e^x=\sum(x^n/n!)$, $x$ is the guage function. If we write $\textrm{exp}(e^x)=\sum(e^{nx}/n!)$, $e^x$ is the gauge function~\cite{VanDykeChapterGauge,Leal}}) may be developed about $\omega = 1$ (i.e. $\eta = 0$)\cite{FlatWallSakiadisPaper}. Note that all constants, including the constant $C$, may be determined via an algorithm independent of any numerical information. The reader is referred to Naghshineh et al.~\cite{FlatWallSakiadisPaper} for a review of literature relevant to the Newtonian Sakiadis problem. 

The objective of this work is to extend the approach of Naghshineh et al. to obtain an analytical solution of the Sakiadis boundary layer for Ostwald-de Waele power law fluids, i.e. the solution of the system~(\ref{eq:ODESystem_NNS}) for $\alpha \neq 1$. This is especially relevant, as many fluids used in thin-film coating exhibit shear thinning character~\cite{Weinstein2004}, for which $\alpha \in (0,1)$. In this parameter range, note that the power law model~(\ref{eq:FoxEquations_NNS}c) is deficient in that it limits to an infinite viscosity, $\mu$, in~(\ref{eq:FoxEquations_NNS}c), as the rate of strain approaches zero ($\partial u/\partial y \to 0$); nevertheless, as the shear stress, $\tau_{xy}$, remains finite in this limit, reasonable predictions may still be made, such as in pipe or slot flows\cite{Hassager}. The power-law model is often used to describe flows for shear thinning behaviour due to its simplicity. However, in this paper, we demonstrate that there is a restricted range of $\alpha$ values ($0.5 < \alpha \leq 1$) for which a solution to the power-law non-Newtonian Sakiadis problem given by~(\ref{eq:ODESystem_NNS}) exists. This is a mathematical restriction of the power-law itself, which is an approximation to the true behavior of a shear thinning fluids~\cite{SchweitzerBook,CoatingPaper}. For fluids that exhibit a power-law dependence satisfying $\alpha < 0.5$ or $\alpha > 1$ over a region of shear rate, calculations must be done with more sophisticated viscosity dependences such as the Carreau Model~\cite{SchweitzerBook, CoatingPaper}. When these models are used in place of $\mu$ in~(\ref{eq:FoxEquations_NNS}c), a similarity variable cannot be identified, and the full PDE system governing boundary layer flows must be solved~\cite{CarreauModel}. 

This paper is organized as follows. In Sec.~\ref{sec:DivergentSection_NNS}, we first examine the solution to the system~(\ref{eq:ODESystem_NNS}) via a standard power series expansion about $\eta = 0$, and find that it is divergent, as was found for Newtonian fluids. In Sec.~\ref{sec:ConvergentSection_NNS}, we then consider the asymptotic solution of the system~(\ref{eq:ODESystem_NNS}) as $\eta \to \infty$, and use it to motivate a Taylor series expansion in terms of an alternative gauge function, as was done for the Newtonian solution~(\ref{eq:Solution_NS}) discussed above. By judiciously choosing the location of the expansion point, we are able to obtain a convergent series solution for power law exponents lying in the range of approximately $0.74\leq \alpha < 1$. For power laws corresponding to $\alpha < 0.74$, the gauge function becomes ill-defined over the full domain. As a result, in Sec.~\ref{sec:Approximant_NNS} an accurate approximate solution is obtained using the method of asymptotic approximants~\cite{Barlow:2017}. Here, the two necessary constants are determined a priori using a numerical shooting method. Note that this approach is taken in the solution of the Falkner-Skan boundary layer equations in prior works~\cite{Cebeci,FS2020}. The utility of the analytical forms are demonstrated in Sec.~\ref{sec:StreamlinesPlots_NNS}, by the ease with which streamlines and the velocity field may be extracted. Concluding comments are provided in Sec.~\ref{sec:Conclusion_NNP}. Formulae used to manipulate the nonlinear series expansion in this study are provided in Appendix~\ref{sec:JCP&Cauchy}. The  shooting algorithm used to solve the non-Newtonian Sakiadis problem~(\ref{eq:ODESystem_NNS}) numerically is provided in Appendix \ref{sec:ShootingMethod}, and relevant constants for the presented power series and approximant are provided in Appendix~\ref{sec:AsymptoticConstants_Table}. Appendix~\ref{sec:PredictionAlgorithm_Appendix} includes the algorithm used to predict the same constants via the convergent power series solution itself.
\section{Power series solution \label{sec:DivergentSection_NNS}}
For this section and the next, we solve the ODE~(\ref{eq:ODESystem_NNS}), which arises after similarity transform. As such, the physics of the original system~(\ref{eq:FoxEquations_NNS}) is obscured in the mathematical solution. In Sec.~\ref{sec:StreamlinesPlots_NNS}, we demonstrate the ease with which streamlines and velocity fields may be extracted from the solution we provide, and in doing so provide solution results in the physical domain. 

A power series solution to the ODE~(\ref{eq:ODESystem_NNS}) can be obtained through standard means using JCP Miller's formula~\cite{Henrici} and Cauchy's product rule~\cite{Churchill} (see Appendixes~\ref{sec:JCP} and \ref{sec:Cauchy}, respectively) to re-order nonlinear terms in powers of $\eta$; the series expansion is
\begin{subequations}
\begin{equation}
    f=\sum_{n=0}^\infty a_n \eta^n,~~|\eta|<\eta_s(\alpha),
    \label{eq:SeriesSolu_S}
\end{equation}
\begin{equation}
     a_{n+3} =  \frac{ \displaystyle \sum_{j=0}^n b_{j}~a_{n-j}}{\alpha(\alpha+1)(n+3)(n+2)(n+1)},~~n\ge0,
\end{equation}
\begin{equation}
    b_{n>0} =\frac{1}{2na_2} \sum_{j=1}^n(3j-\alpha j-n)(j+2)(j+1)a_{j+2}b_{n-j}~~,~~ b_0 = (-2a_2)^{2-\alpha},
\end{equation}
\begin{equation}
    a_0 = 0~~,~~a_1 = 1~~,~~\textrm{and} \  a_2 = \kappa/2,
\end{equation}
\label{eq:DivSeriesSolu_NNS}
\end{subequations}
where $\eta_s(\alpha)$ is a finite radius of convergence. In~(\ref{eq:DivSeriesSolu_NNS}d), the quantity $\kappa$ is directly related to the wall shear stress in the boundary layer flow, typically referred to as the "wall shear" parameter~\cite{Cortell,Fazio2015}, and is defined as
\begin{equation}
    \kappa = f''(0).
    \label{eq:kappa}
\end{equation} 
 The quantity $\kappa$ in~(\ref{eq:kappa}) is a function of $\alpha$, and is not known a priori; it is typically determined numerically. Alternatively, $\kappa$ can be calculated algorithmically as shown in Sec.~\ref{sec:PredictionAlgorithm} of this paper as an extension of the technique developed for the Newtonian Sakiadis problem\cite{FlatWallSakiadisPaper}.

\begin{figure}[htbp]
  \centering
  \includegraphics[width=11cm]{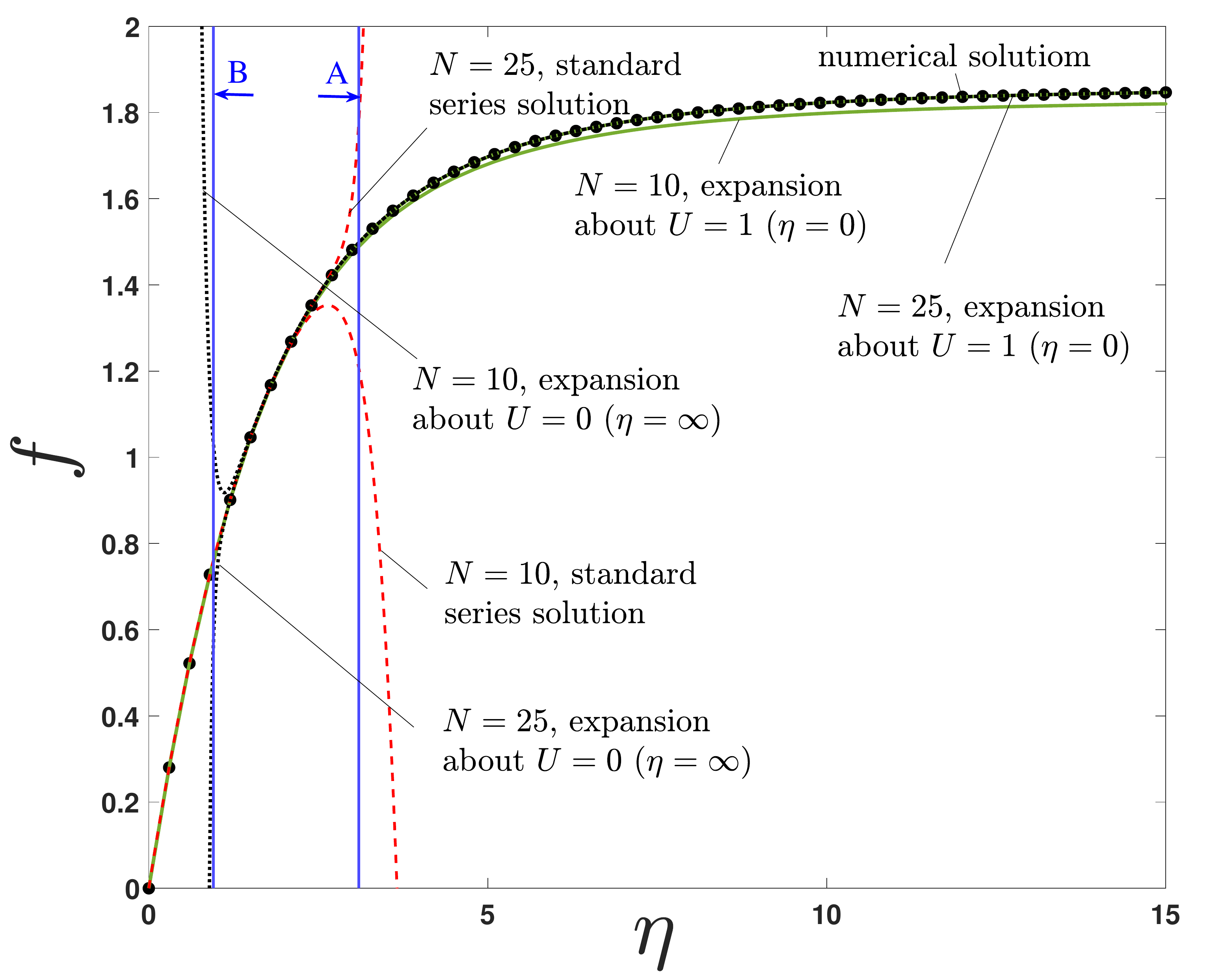}
  \caption{The solution to~(\ref{eq:ODESystem_NNS}) is shown for $\alpha = 0.8$. The numerical solution (Appendix~\ref{sec:ShootingMethod}) with $L=11000$ (black dots) is compared against the $N$-term truncations of the power series solution~(\ref{eq:DivSeriesSolu_NNS}) (dashed curves), transformed series solution (expansion about $\eta = \infty$) (\ref{eq:UTransformedSeriesSolu_NNS}) (dotted curves), and the  transformed series solution (expansion about $\eta = 0$) (\ref{eq:UtTransformedSeriesSolu_NNS}) (solid curves) for $N=10$ and $N=25$. In regions of the plot where a given dashed or dotted curve is not clearly seen, the curves agree with the numerical results. The rightmost vertical line (marked by arrow A) shows the radius of convergence $\eta_s \approx 3.09$ of the standard series solution~(\ref{eq:DivSeriesSolu_NNS}), and the leftmost vertical line (marked by arrow B) shows the radius of convergence $\eta_s \approx 0.95$ of the transformed series solution~(\ref{eq:UTransformedSeriesSolu_NNS}). For $\alpha = 0.8$, the numerically obtained values of the constants $\kappa$, $C$, and $E$ (defined in Sec.\ref{sec:ConvergentSection_NNS}) used in producing the figure are given in Appendix~\ref{sec:NumericalValues_Table}.} 
  \label{fig:ResummationSolu_NNS}
\end{figure}

\begin{figure}[htbp]
  \centering
  \includegraphics[width=10cm]{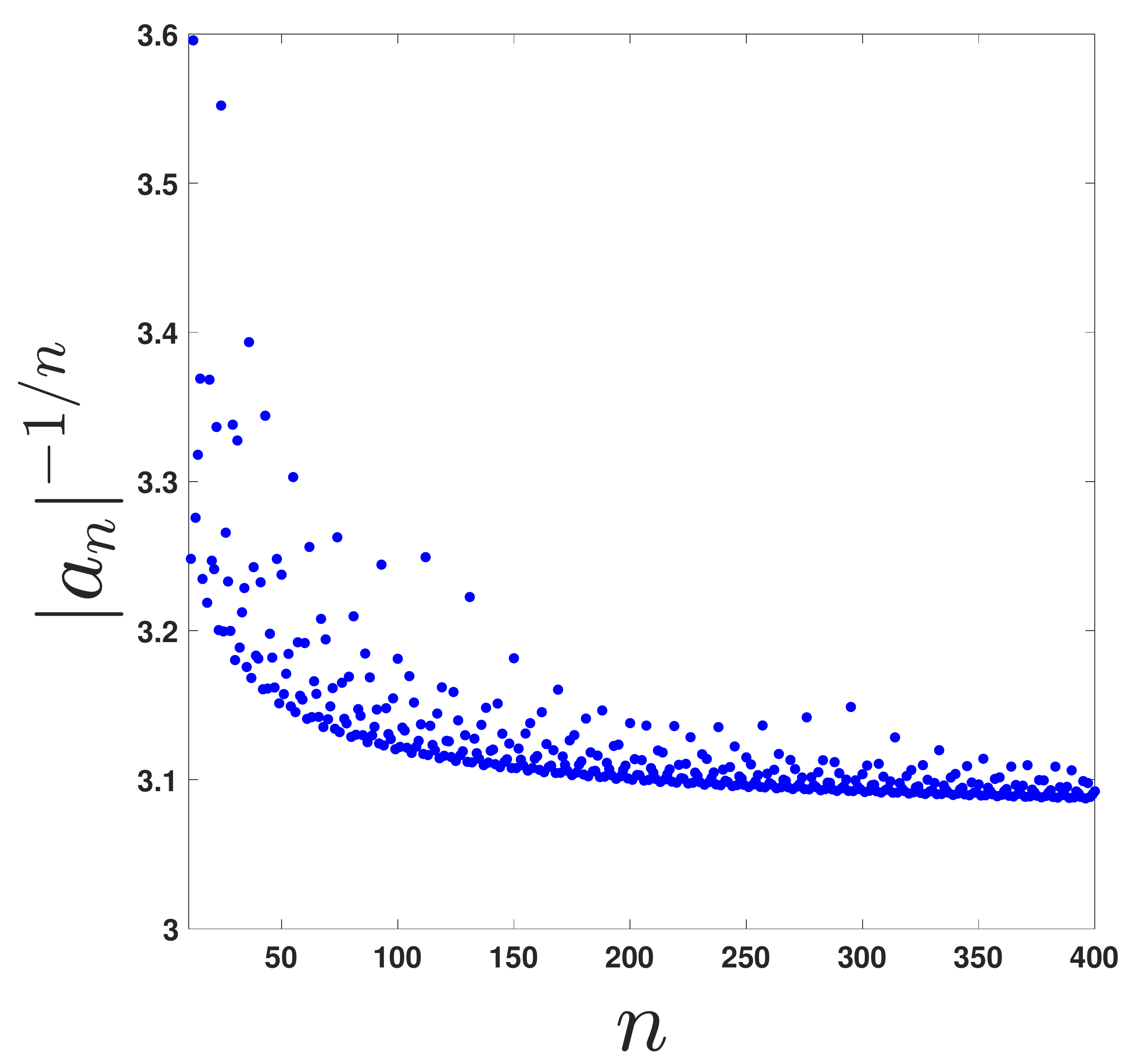}
  \caption{Root test for~(\ref{eq:DivSeriesSolu_NNS}) is shown for $\alpha = 0.8$, indicating a radius of convergence ($y$-axis) of $\eta_s \approx 3.09$. This is consistent with the divergent behavior observed in Fig.~\ref{fig:ResummationSolu_NNS}.}
  \label{fig:Domb-Sykes_stdSeriesSolu_NNS}
\end{figure}

Figure~\ref{fig:ResummationSolu_NNS} provides a comparison between the power series solution~(\ref{eq:DivSeriesSolu_NNS}) and the numerical solution to~(\ref{eq:ODESystem_NNS}) for $\alpha = 0.8$. The numerical solution is obtained using a shooting method (see Appendix~\ref{sec:ShootingMethod}) to recast system~(\ref{eq:ODESystem_NNS}) as a boundary value problem on a finite domain length $L$, where the condition~(\ref{eq:f'(inf)BC_SP_NNS}) is replaced with $df/d\eta= 0$ at $\eta = L$; the length $L$ is chosen such that doubling its size leads to difference in the predictions of $\kappa$ and $C = f(L)$ of $O(10^{-15})$ and $O(10^{-10})$ when $\alpha=0.8$. The constants $\kappa$ and $C$ are defined in~(\ref{eq:kappa}) and (\ref{eq:AsymsptoticC_NNS}), respectively. As shown in the Fig.~\ref{fig:ResummationSolu_NNS}, the power series solution~(\ref{eq:DivSeriesSolu_NNS}) diverges within the physical domain. The rightmost vertical line marked by arrow $A$ shows the radius of convergence of the power series solution~(\ref{eq:DivSeriesSolu_NNS}) given by $\eta_s(0.8) \approx 3.09$ , which is confirmed via a numerical root test \footnote{Some of the coefficients $a_n$ in~(\ref{eq:DivSeriesSolu_NNS}) are zero; hence, we use root test instead of ratio test.}, as shown in Fig.~\ref{fig:Domb-Sykes_stdSeriesSolu_NNS}.

\section{Asymptotically motivated Gauge Function and Expansions \label{sec:ConvergentSection_NNS}}
\subsection{Asymptotic Behavior as $\eta \rightarrow \infty$} \label{sec:Transformations}
Similar to the approach taken for Newtonian fluids~\cite{FlatWallSakiadisPaper}, we use the method of dominant balance~\cite{Bender} to determine the asymptotic behavior of $f$ as $\eta \to \infty$. This behavior motivates the use of a gauge function that ultimately leads to a convergent series expansion. To proceed, we write the solution of~(\ref{eq:ODESystem_NNS}a) as
\begin{subequations}
\begin{equation}
    f \sim C + h(\eta),~\textrm{with} \  h \to 0 ~\textrm{as} \  \eta \to \infty,
    \label{eq:AsympSolu_NNS}
\end{equation}
where $C$ is the asymptotic constant described in (\ref{eq:AsymsptoticC_NNS}), and $h(\eta)$ is a function to be determined. The form~(\ref{eq:AsympSolu_NNS}) is substituted in~(\ref{eq:ODE_NNS}) to obtain
\begin{equation}
    \alpha(\alpha+1) h''' \sim (C+h)(-h'')^{(2-\alpha)}  ~\textrm{as} \  \eta \to \infty,
    \label{Asymp_ODE_2order_NNS}
\end{equation}
where the primes denote derivatives of $h$ with respect to $\eta$. Equation~(\ref{Asymp_ODE_2order_NNS}) may be simplified by noting that $h$ is subdominant to $C$ as $\eta \to \infty$, and thus
\begin{equation}
    \alpha(\alpha+1) h'''  \sim C(-h'')^{(2-\alpha)} ~\textrm{as}  \  \eta \to \infty.
\end{equation}
The above equation can be integrated once to obtain 
\begin{equation}
    h''\sim -\left[ E+\frac{C(1-\alpha)}{\alpha(\alpha+1)}\eta \right]^{\frac{1}{\alpha-1}} ~\textrm{as} \  \eta \to \infty,
    \label{eq:h''_NNS}
\end{equation}
where $E$ is the constant of integration to be determined. Integrating~(\ref{eq:h''_NNS}) twice, and applying the boundary condition~(\ref{eq:f'(inf)BC_SP_NNS}), the solution of~(\ref{eq:h''_NNS}) is 
\begin{equation}
    h \sim \displaystyle\frac{\alpha(\alpha+1)^2E^\frac{2\alpha-1}{\alpha-1}}{-C^2(2\alpha-1)}\left[ 1+\frac{C(1-\alpha)}{\alpha(\alpha+1)E}\eta\right]^{\frac{2\alpha-1}{\alpha-1}}~~,~~ 
    0.5 < \alpha <1~,~\textrm{as} \  \eta \to \infty,
    \label{eq:h_NNS}
\end{equation}
and thus from (\ref{eq:h_NNS}), we obtain
\begin{equation}
    f \sim C+ \displaystyle\frac{\alpha(\alpha+1)^2E^\frac{2\alpha-1}{\alpha-1}}{-C^2(2\alpha-1)}\left[ 1+\frac{C(1-\alpha)}{\alpha(\alpha+1)E}\eta\right]^{\frac{2\alpha-1}{\alpha-1}}~~,~~
    0.5 < \alpha <1~,~\textrm{as} \  \eta \to \infty.
    \label{eq:AsymptoticSoluFinal_NNS}
\end{equation}
\label{eq:AsyAsymptoticSolu_All}
\end{subequations}
By inspection, we see that~(\ref{eq:AsymptoticSoluFinal_NNS}) approaches $C$ as $\eta \to \infty$ only when $0.5 < \alpha <1$, and thus condition~(\ref{eq:f'(inf)BC_SP_NNS}) can only be satisfied in this range. Consequently, system~(\ref{eq:ODESystem_NNS}) is only valid for $0.5 < \alpha \leq 1$ ($\alpha = 1$ for Newtonian fluids). Note that Fox et al.~\cite{Fox} incorrectly indicate that the solution to system~(\ref{eq:ODESystem_NNS}) exists when $0 < \alpha <0.5$. Additionally, although such solutions may be obtained to the finite-domain approximation to system~(\ref{eq:ODESystem_NNS}) where~(\ref{eq:f'(inf)BC_SP_NNS}) is replaced with $f'(L) = 0$, these solutions do not converge to an infinite domain solution as $L \to \infty$; note that Pop et al.~\cite{Pop} incorrectly claim that the solution exists for $\alpha > 1$. The reader is referred to comments made in Sec.~\ref{sec:Intro_NNS} regarding the physical significance of this restriction.
\subsection{Construction of a Convergent Power Series Solution} \label{sec:ConvResummation}
To overcome the convergence limitation of the power series solution~(\ref{eq:DivSeriesSolu_NNS}), we follow the approach of Naghshineh et al.~\cite{FlatWallSakiadisPaper} for the Newtonian problem (see Sec.~\ref{sec:Intro_NNS}). Here, we propose the following variable transformation, inspired by the asymptotic expansion~(\ref{eq:AsymptoticSoluFinal_NNS}), whose utility is validated in what follows. We write

\begin{subequations}
\begin{equation}
    U(\eta) =  \left[ 1+\mathcal{A}\eta\right] ^\lambda,
\end{equation}
\begin{equation}
    f(\eta) = F(U(\eta)),
\end{equation}   
\begin{equation}
    \mathcal{A} = \frac{C(1-\alpha)}{\alpha(\alpha+1)E}~~,~~\lambda = \frac{2\alpha-1}{\alpha-1}.
    \label{eq:a_and_lambda_NNS}
\end{equation}
\label{eq:UTransformation_NNS}
\end{subequations}
The transformation~(\ref{eq:UTransformation_NNS}) maps $\eta \in [0,\infty)$ to $U \in (0,1]$ when $0.5 < \alpha <1$. 

It is worth noting that the non-Newtonian transformation in~(\ref{eq:UTransformation_NNS}a), reduces to the Newtonian transformation given by~(\ref{eq:Solution_NS}), as $\alpha \to 1$, since
\begin{equation}
    \lim_{\alpha \rightarrow 1}\left[ 1+\mathcal{A}\eta\right] ^\lambda = e^{\displaystyle -C\eta/2E},
    \label{eq:limit_NNS}
\end{equation}
when $E = 1$; we have indeed verified numerically that $E \to 1$ as $\alpha \to 1$ (see Table~\ref{tab:Table_NumE}). Substituting~(\ref{eq:UTransformation_NNS}) into~(\ref{eq:ODE_NNS}), applying the chain rule, and rearranging terms, we obtain the transformed ODE 
\begin{equation*}
    k_1 U^{\left(3-\frac{3}{\lambda}\right)}F''' + k_2U^{\left(2-\frac{3}{\lambda}\right)}F''+k_3U^{\left(1-\frac{3}{\lambda}\right)}F' - 
    \left\{k_4U^{\left(2-\frac{2}{\lambda}\right)}F''+k_5U^{\left(1-\frac{2}{\lambda}\right)}F'   \right\}^{2-\alpha}F = 0, 
    \label{eq:TransformedODE_Unsimplified_NNS}
\end{equation*}
where the primes denote derivatives of $F$ with respect to $U$. After multiplying the above by $U^{\left(\frac{3}{\lambda}-1\right)}$, and rearranging the ODE such that the highest derivative is on the left side of the equation, we obtain
\begin{subequations}
\begin{equation}
    k_1 U^2F''' = -k_2UF''-k_3F' + \left\{k_4UF''+k_5F'\right\}^{2-\alpha}F,
    \label{eq:TransformedODE_Simplified_NNS}
\end{equation}
where
\begin{align}
    k_1=&\alpha(\alpha+1)\mathcal{A}^3\lambda^3,
    \label{eq:k1}\\
    k_2=&3\alpha(\alpha+1)\mathcal{A}^3\lambda^2(\lambda-1),
    \label{eq:k2}\\
    k_3=&3\alpha(\alpha+1)\mathcal{A}^3\lambda(\lambda-1)(\lambda-2),
    \label{eq:k3}\\
    k_4=&-\mathcal{A}^2\lambda^2,
    \label{eq:k4}\\
    k_5=&-\mathcal{A}^2\lambda(\lambda-1),
    \label{eq:k5}
\end{align}
and the expressions for $\mathcal{A}$ and $\lambda$ are defined in (\ref{eq:a_and_lambda_NNS}). The boundary conditions at $\eta = 0$ from~(\ref{eq:f(0)BC_SP_NNS}) and~(\ref{eq:f'(0)BC_SP_NNS}), corresponding to $U = 1$, become
\begin{equation}
    F(1) = 0,
\end{equation}
\begin{equation}
    F'(1) = \frac{1}{\mathcal{A} \lambda},
\end{equation}
\begin{equation}
    F''(1) = \frac{\kappa}{\mathcal{A}^2 \lambda^2} - \frac{\lambda - 1}{\mathcal{A} \lambda^2},
\end{equation}
\label{eq:Transformed1BCs_NNS}
\end{subequations}
where the boundary condition~(\ref{eq:kappa}) is used to obtain $F''(1)$ in (\ref{eq:Transformed1BCs_NNS}i). 

We next assume a solution to~(\ref{eq:TransformedODE_Simplified_NNS}) of the form \begin{equation}
    F(U) = \sum_{n=0}^\infty {A}_n U^n,~|U|<U_s(\alpha)~,~E>0,
    \label{eq:AssumedF(U)}
\end{equation} 
where $U_s(\alpha)$ is the radius of convergence. The $E>0$ restriction allows for the use of~(\ref{eq:AsyAsymptoticSolu_All}) without introducing branch point singularities into the $U$ domain of interest. Series~(\ref{eq:AssumedF(U)}) is readily differentiated term-by-term to compute $F'$, $F''$, and $F'''$. After employing JCP Miller's formula~\cite{Henrici} and Cauchy's product rule~\cite{Churchill} (see Appendixes~\ref{sec:JCP} and~\ref{sec:Cauchy}, respectively) to re-order the nonlinear terms in~(\ref{eq:TransformedODE_Simplified_NNS}), the ODE in~(\ref{eq:TransformedODE_Simplified_NNS}) becomes 

\begin{widetext}
\begin{multline}
    \sum_{n=2}^\infty k_1(n+1)n(n-1)A_{n+1}U^n =  
    (-k_3A_1+A_0d_0)+ (-2k_2A_2-2k_3A_2+A_0d_1+A_1d_0)U + \\
    \sum_{n=2}^\infty \left(-k_2(n+1)nA_{n+1} - K_3(n+1) + A_0\tilde{d}_n + \frac{(2-\alpha)A_0d_0}{c_0}(n+1)(k_4n+k5)A_{n+1} +\tilde{e}_n\right)U^n, 
    \label{eq:TransformedExpandedForm_NNS}
\end{multline}
\end{widetext}
where
\begin{subequations}
\begin{equation}
    \tilde{d}_n =\frac{1}{nc_0}\sum_{j=1}^{n-1}(3j-\alpha j -n)c_jd_{n-j}~~, ~~\tilde{e}_n=\sum_{j=1}^{n}A_jd_{n-j},
    \label{eq:d_and_e_NNS}
\end{equation}
\begin{equation}
    d_{n>0} =\frac{1}{nc_0}\sum_{j=1}^{n}(3j-\alpha j -n)c_jd_{n-j}~~,~~d_0 = (c_0)^{(2-\alpha)},
\end{equation}
\begin{equation}
    c_{n>0}=(n+1)A_{n+1}(k_4n+k_5)~~,~~c_0=k_5A_1.
\end{equation}

Using the asymptotic solution~(\ref{eq:AsymptoticSoluFinal_NNS}), we enforce
\begin{equation}
    A_0 = C.
\label{eq:A0_NNS}
\end{equation}
Equating constant terms on both sides of~(\ref{eq:TransformedExpandedForm_NNS}) leads to 
\begin{equation}
   {A}_1= \left[\frac{k_3}{A_0(k_5)^{2-\alpha}}\right]^{\frac{1}{1-\alpha}},
   \label{eq:A1_NNS}
\end{equation}
and equating $U^1$ terms on both sides of~(\ref{eq:TransformedExpandedForm_NNS}) leads to  
\begin{equation}
    A_2=\frac{A_1d_0}{2k_2+2k_3-\frac{2(2-\alpha)A_0d_0}{c_0}(n+1)(k_4n+k5)}.
    \label{eq:A2_NNS}
\end{equation}
We equate like-terms in~(\ref{eq:TransformedExpandedForm_NNS}) to obtain the coefficients $A_{n+1}$, and for $n\geq 2$, we obtain the recurrence relation
\begin{widetext}
\begin{equation}
    A_{n+1}=\frac{A_0\tilde{d}_n + \tilde{e}_n}{K_1(n+1)n(n-1)+k_2(n+1)n+k_3(n+1)-\frac{(2-\alpha)A_0d_0}{c_0}(n+1)(k_4n+k5)}~, ~n\geq 2,
     \label{eq:TransforedODECoeff_NNS}
\end{equation}
\end{widetext}
where $\tilde{d}_n$ and $\tilde{e}_n$ are defined in~(\ref{eq:d_and_e_NNS}).
Transforming back to $f(\eta)$ space via~(\ref{eq:UTransformation_NNS}), our expansion about $U=0$ (i.e. $\eta = \infty$) is
\begin{equation}
    f(\eta) =\sum_{n=1}^\infty A_{n} \left[ 1+\mathcal{A}\eta\right] ^\lambda.
\end{equation}
\label{eq:UTransformedSeriesSolu_NNS}
\end{subequations}

\begin{figure}[htbp]
  \centering
  \includegraphics[width=10cm]{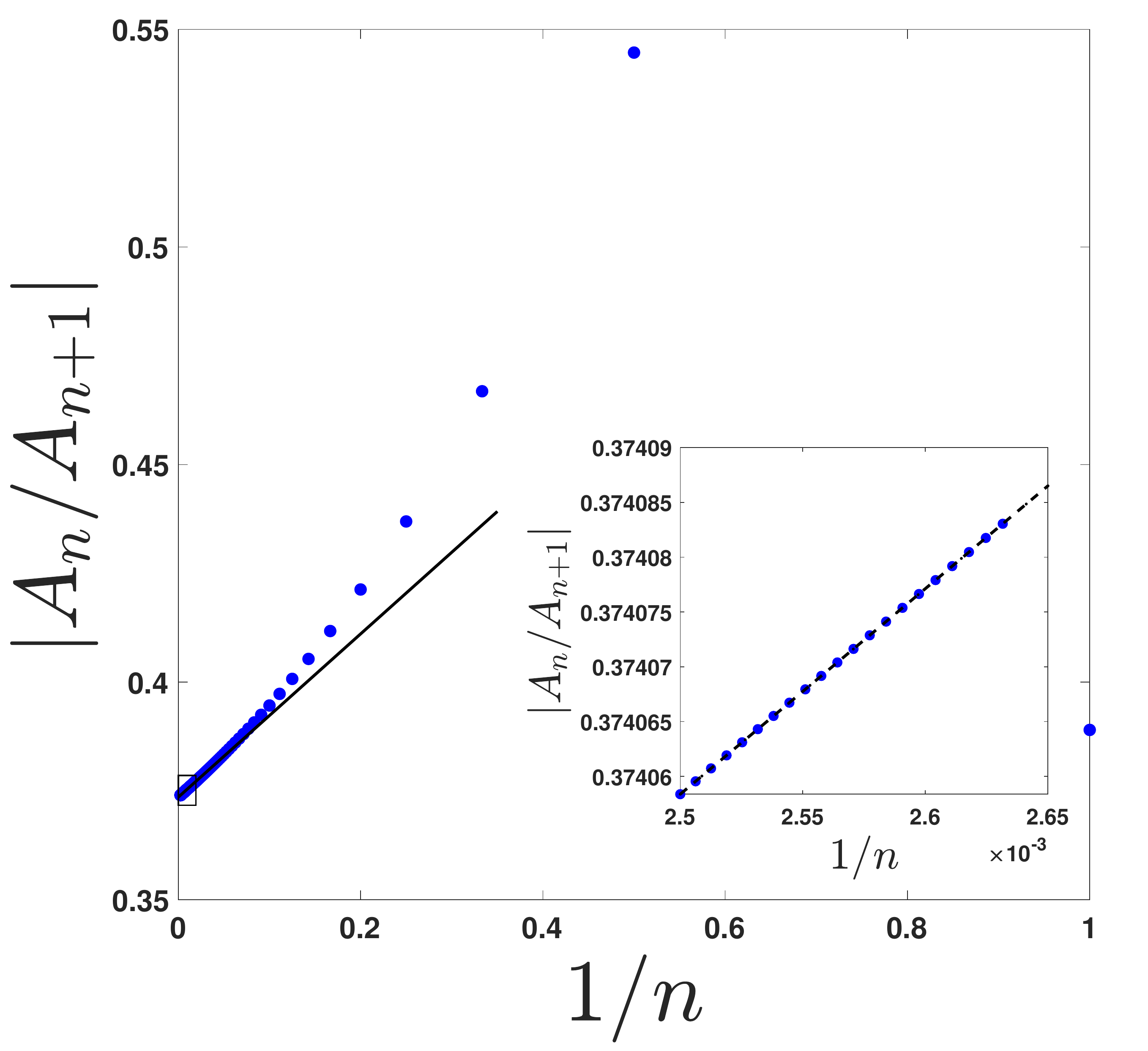}
  \caption{Domb-Sykes plot for~(\ref{eq:UTransformedSeriesSolu_NNS}) with $\alpha = 0.8$, Here, the intercept for $1/n = 0$ yields the numerical radius of convergence ($y$-axis) $U_s \approx 0.37$, in agreement with a radius of convergence in the original domain $\eta_s \approx 0.95$ through~(\ref{eq:UTransformation_NNS}a). This is consistent with the divergent behavior observed in Fig.~\ref{fig:ResummationSolu_NNS}.} 
  \label{fig:Domb-Sykes_UTransformed_NNS}
\end{figure}

Although our ultimate goal is to have a self-contained solution that is not dependent on numerically determined parameters, at this stage we use the numerical values of $\kappa$, $C$, and $E$ in~(\ref{eq:UTransformedSeriesSolu_NNS}) (see Table~\ref{tab:Table_NumValues_Alpha0.8}) to assess the efficacy of~(\ref{eq:UTransformedSeriesSolu_NNS}). To determine the numerical value of $E$, we solve~(\ref{eq:h''_NNS}) for $E$ as
\begin{equation}
    E = (-h'')^{\alpha-1}-\frac{C(1-\alpha)}{\alpha(\alpha+1)}\eta~~\textrm{as} \  \eta \to \infty,
    \label{eq:E_NNS}
\end{equation}
where the $\eta \to \infty$ condition is approximated in a numerical solution (of domain length $L$) by replacing $\eta$ with $L$, $C$ with $h(L)$, and $h''$ with $h''(L)$.
Figure~\ref{fig:ResummationSolu_NNS} shows that the transformed series solution~(\ref{eq:UTransformedSeriesSolu_NNS}) (dotted curves) matches the numerical solution as $\eta \to \infty$, and the standard power series solution~(\ref{eq:DivSeriesSolu_NNS}) (dashed curves) matches the numerical solution as $\eta \to 0$, as expected. It is apparent here that the power series solution~(\ref{eq:UTransformedSeriesSolu_NNS}) diverges as $\eta \rightarrow 0$. Since the coefficients of~(\ref{eq:UTransformedSeriesSolu_NNS}) alternate in sign, the closest singularity lies along the negative real $U$ axis\cite{VanDyke}, i.e. outside the physical domain. The vertical solid line (marked by arrow B) in Fig.~\ref{fig:ResummationSolu_NNS} shows the radius of convergence of the series solution~(\ref{eq:UTransformedSeriesSolu_NNS}). This radius is confirmed via a numerical ratio test in the form of a Domb-Skyes plot\cite{VanDyke}, shown in Fig.~\ref{fig:Domb-Sykes_UTransformed_NNS} as a plot of the relevant coefficient ratio vs. $1/n$. As the curve is linear in $1/n$ for large $n$, the radius of convergence is identified as the $y$-intercept. 

Naghshineh et al.~\cite{FlatWallSakiadisPaper} show that the radius of convergence for the \textit {Newtonian} Sakiadis problem can be increased by changing the expansion point of the power series solution to the transformed ODE. Inspired from their work, we change the expansion point in the power series solution to~(\ref{eq:Transformed1BCs_NNS}) to $U=1$, corresponding to $\eta = 0$. Using the same procedures employed above to obtain the series about $U=0$, the series about $U = 1$ is defined as
\begin{subequations}
\begin{equation}
    F(U) = \sum_{n=0}^\infty \hat{A}_n (U-1)^n,~E>0,
\end{equation}
where
\begin{equation}
    \hat{A}_0 = 0,
\end{equation}
\begin{equation}
    \hat{A}_1 = \frac{1}{\mathcal{A}\lambda},
\end{equation}
\begin{equation}
    \hat{A}_2 = \frac{1}{2} \left(\frac{\kappa}{\mathcal{A}^2 \lambda^2} - \frac{\lambda - 1}{\mathcal{A} \lambda^2} \right),
\end{equation}
\begin{equation}
    \hat{A}_3 = \frac{1}{6k_1}\left(  -2k_2\hat{A}2 - k_3 \hat{A}_1 \right),
\end{equation}
\begin{equation}
    \hat{A}_4 = \frac{1}{24k_1}\left(  -2k_2\hat{A}2 - 6k_2 \hat{A}_3 - 2k_3 \hat{A}_2 + \hat{A}_1 \hat{d}_0-12k_1 \hat{A}_3\right).
\end{equation}

Following the same approach as was employed earlier, we obtain
\begin{multline}
    \hat{A}_{n+3} = \frac{\left\{-k_2(n+1)n - k_3(n+1) - k_1(n+1)n(n-1)\right\}\hat{A}_{n+1}}{k_1(n+3)(n+2)(n+1)} +  \\
    \frac{\left\{   -k_2(n+2)(n+1) - 2k_1(n+2)(n+1)n \right\}\hat{A}_{n+2} + \hat{e}_n}{k_1(n+3)(n+2)(n+1)}~,~n \geq 2,
\end{multline}
with
\begin{equation}
     \hat{d}_n =\frac{1}{n\hat{c}_0}\sum_{j=1}^{n}(3j-\alpha j -n)\hat{c}_j\hat{d}_{n-j}~~,~~ \hat{d}_0=(\hat{c}_0)^{(2-\alpha)}\\
\end{equation}
\begin{equation}
    \hat{e}_n=\sum_{j=0}^{n}\hat{A}_j\hat{d}_{n-j}
\end{equation}
\begin{equation}
    \hat{c}_{n>0}=(k_4n+k_5)(n+1)\hat{A}_{n+1}+k_4(n+2)(n+1)\hat{A}_{n+2} ~~,~~\hat{c}_0=2k_4\hat{A}_2+k_5\hat{A}_1.
\end{equation}
The constants $k_1$ - $k_5$ are defined in~(\ref{eq:k1}) -~(\ref{eq:k5}). 
\label{eq:UtTransformedSeriesSolu_NNS}
\end{subequations}

\begin{figure} [htbp]
\centering
    \centering
    \subfloat{{\includegraphics[width=8.5cm]{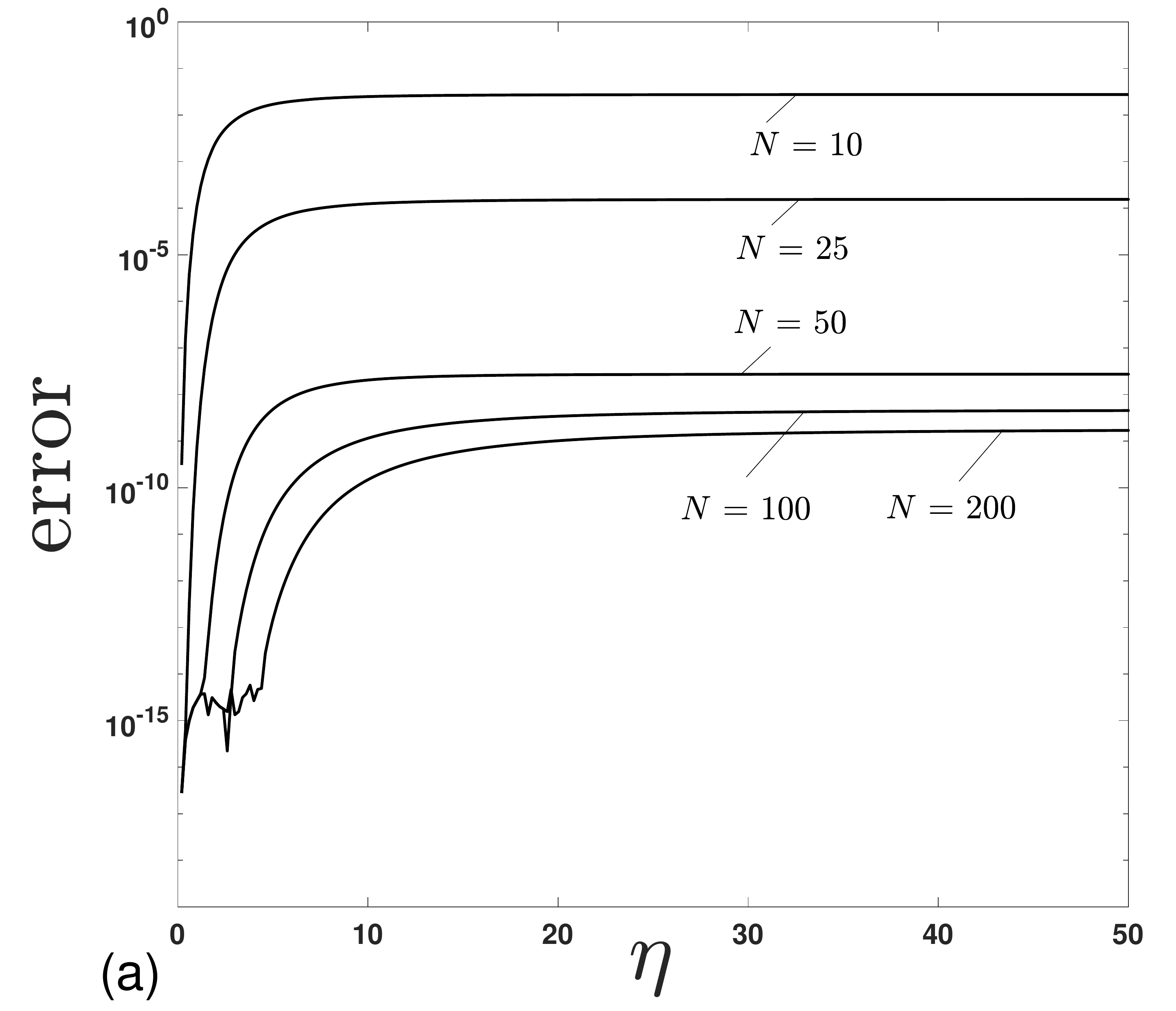} }}
    \subfloat{{\includegraphics[width=8.45cm]{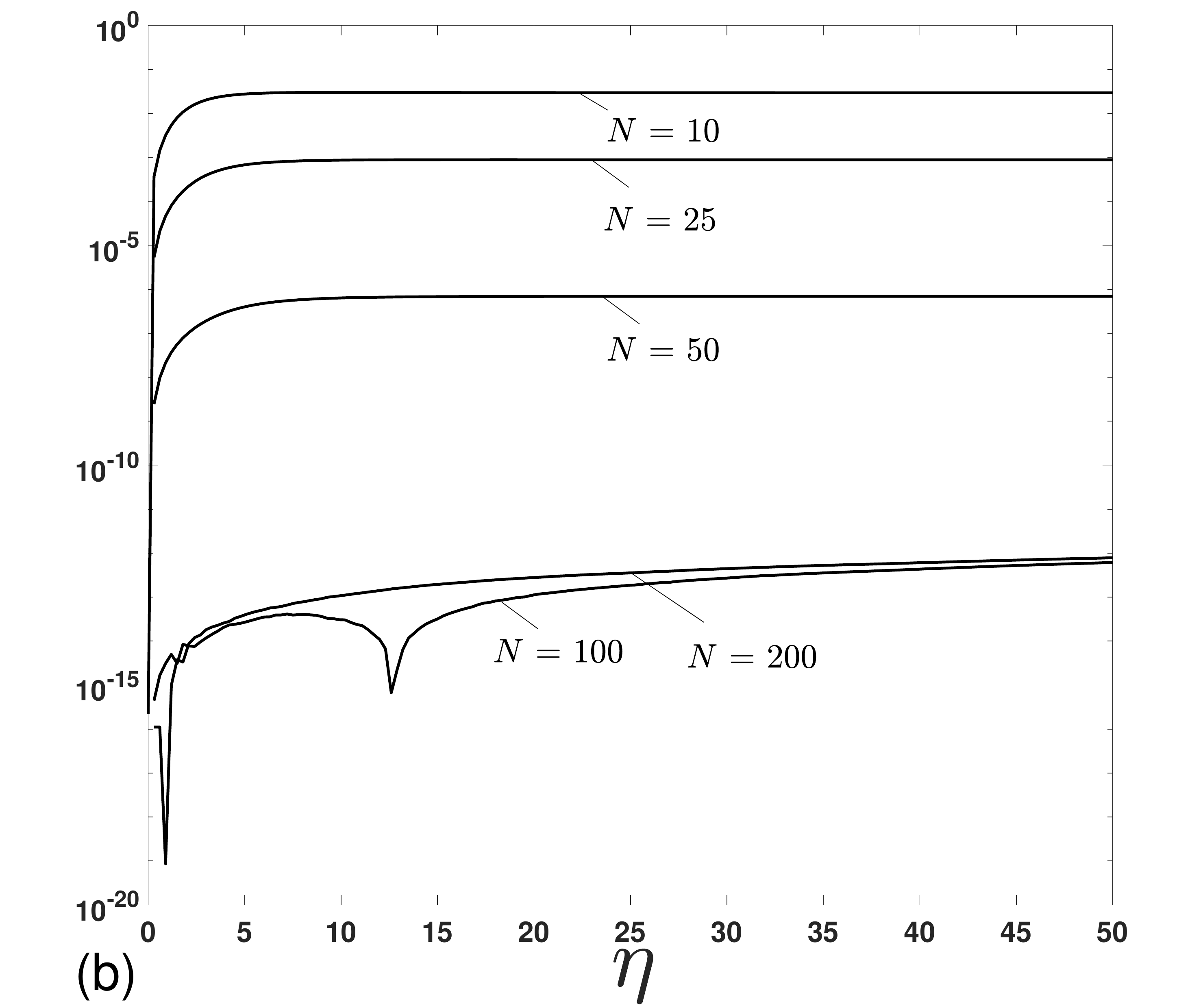} }}
    \caption{Absolute error (the absolute value of the difference) between $N$-term truncations of the convergent series solution~(\ref{eq:UtTransformedSeriesSolu_NNS}) and the numerical solution (over the domain $\eta \in [0,L]$ with $L=11000$) for $\alpha = 0.8$, plotted versus $\eta$, using $\kappa$, $C$, and $E$ values (a) generated by the numerical solution, and (b) predicted by equations system~(\ref{eq:EquationSystem_NNS}), discussed in Sec.~\ref{sec:PredictionAlgorithm}. }
    \label{fig:ErrorPlots1_NNS}
\end{figure}

\begin{figure} [htbp]
\centering
    \centering
    \subfloat{{\includegraphics[width=10cm]{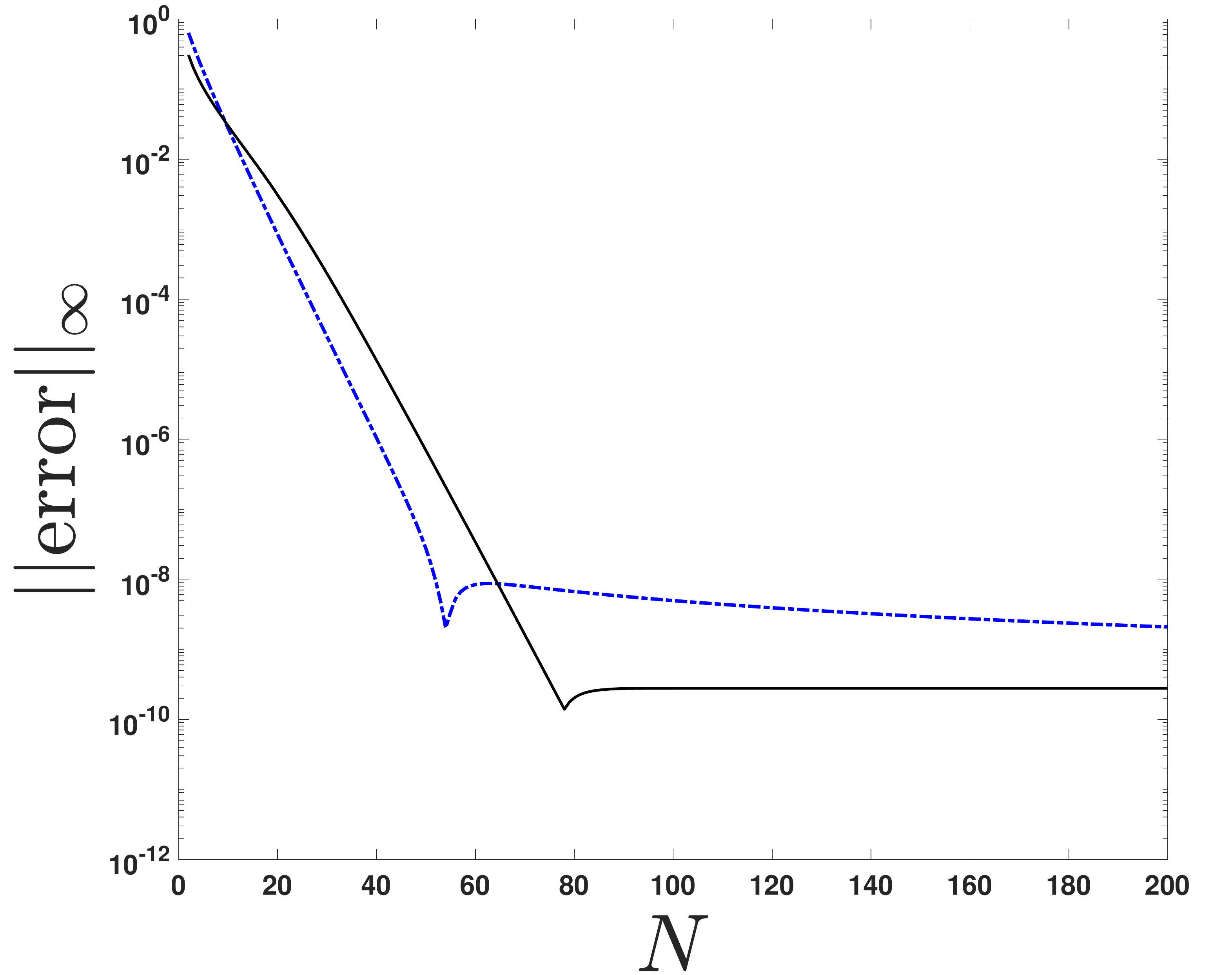} }}
    \caption{The infinity norm (maximum absolute error) between $N$-terms truncations of (\ref{eq:UtTransformedSeriesSolu_NNS}) and the numerical solution (occurring over $\eta \in [0,L]$ with $L=11000$) for $\alpha = 0.8$, plotted versus $N$. The dashed and solid curves correspond to when numerical and algorithmically predicted values (using equations system~(\ref{eq:EquationSystem_NNS}) discussed in Sec.~\ref{sec:PredictionAlgorithm}) of $\kappa$, $C$, and $E$ are used, respectively. Convergence of the numerical and algorithmically predicted values of the constants are reported in respective Tables~\ref{tab:Table_NumValues_Alpha0.8} and~\ref{tab:Table_PredictedValues_Alpha0.8} of Appendix~\ref{sec:NumericalValues_Table}.}
    \label{fig:ErrorPlots2_NNS}
\end{figure}

Figure~\ref{fig:ResummationSolu_NNS} shows that the transformed series solution~(\ref{eq:UtTransformedSeriesSolu_NNS}) (solid curves) matches the numerical solution as $\eta \to \infty$, as well as $\eta \to 0$. Thus, moving the location of the expansion point from $U=0$ to $U=1$ enables a convergent expansion over the whole domain. It should be noted here that this is distinctly different from the Newtonian case\cite{FlatWallSakiadisPaper}, where \textit{both} expansions about $U=0$ and $U=1$ converge--although the latter expansion converges faster. Figure~\ref{fig:ErrorPlots1_NNS}a shows the absolute error (the absolute difference) between $N$-term truncations of the
convergent series solution~(\ref{eq:UtTransformedSeriesSolu_NNS}) and the numerical solution for $\alpha = 0.8$. The numerical values of the constants used to generate this figure are shown in Appendix~\ref{sec:NumericalValues_Table}. Here, we choose to stop at $N=200$ in construction of Fig.~\ref{fig:ErrorPlots1_NNS} because the absolute error is close to machine precision. The dashed curve in Fig.~\ref{fig:ErrorPlots2_NNS} shows the infinity norm (maximum absolute error) between the $N$-term truncation of the series solution~(\ref{eq:UtTransformedSeriesSolu_NNS}) and the numerical solution taken over $\eta \in [0,L]$ with $L=11000$ for $\alpha = 0.8$, using the values of constants generated by the numerical solution (see Appendix~\ref{sec:NumericalValues_Table} for more details). The plateau reached in this figure occurs when further refinements to the power series lead to errors smaller than that of the numerical solution. 

\subsection{Prediction of Unknown Parameters} \label{sec:PredictionAlgorithm}
Thus far, results have been presented where the numerically-obtained values of the constants $\kappa$, $C$, and $E$ have been used. With the aim of making the series solution~(\ref{eq:UtTransformedSeriesSolu_NNS}) independent of the numerical solution, we adapt an algorithm used by Barlow et al.~\cite{Barlow:2017} to predict the values of the constants $\kappa$, $C$, and $E$. In that study, it is sufficient to construct a system of equations for the unknowns by choosing the last $\mathcal{N}$ coefficients of the series solution to be zero, where $\mathcal{N}$ is the number of unknowns ($\mathcal{N}= 3$ in this problem: $\kappa$, $C$, and $E$). For a convergent series, it is most certainly the case that the last coefficient approaches zero as $N \to \infty$, so this assumption is self consistent in the limit. Additionally, from the perspective of the number of equations and unknowns, we need three equations to find $\kappa$, $C$, and $E$. The assumption of series convergence implicit in the equations to solve is validated by convergence of the algorithm itself for increasingly large numbers of series terms. However, in this non-Newtonian extension we find that the three equations ($\hat{A}_N = 0$, $\hat{A}_{N-1} = 0$, and $\hat{A}_{N-2} = 0$) are linearly dependent, as evident by the determinant of the Jacobian. For that reason, we alter the algorithm such that we use one of those equations ($\hat{A}_N = 0$)~(\ref{eq:F3}). For the remaining two equations we adapt the algorithm used by Naghshineh et al.~\cite{FlatWallSakiadisPaper} where conditions are imposed from the side of the domain that is opposite to that of the series' expansion point. The system of equations used here are:
\begin{subequations}
    \begin{equation}
        \left[ \sum_{n=0}^N \hat{A}_n (-1)^n\right]-C = 0,
        \label{eq:F1}
    \end{equation}
    \begin{equation}
        \left[ \sum_{n=0}^N n\hat{A}_n (-1)^{n-1}\right]-A_1 = 0 ,
        \label{eq:F2}
    \end{equation}
    \begin{equation}
        \hat{A}_N = 0,
        \label{eq:F3}
\end{equation} 
\label{eq:EquationSystem_NNS}
\end{subequations}
where $C$, $A_1$, and $\hat{A}_N$ are defined in~(\ref{eq:AsymsptoticC_NNS}), (\ref{eq:A1_NNS}), and~(\ref{eq:UtTransformedSeriesSolu_NNS}g), respectively. Note that~(\ref{eq:F1}) and~(\ref{eq:F2}) correspond to the boundary conditions $F(U=0) = C$ and $F'(U=0) = A_1$. Newton's Method is used to solve the system of equations~(\ref{eq:EquationSystem_NNS}), and the details are provided in Appendix~\ref{sec:PredictionAlgorithm_Appendix}.

Figures~\ref{fig:ErrorPlots1_NNS}b and~\ref{fig:ErrorPlots2_NNS} (solid curve) show typical results of the solution of equations system~(\ref{eq:EquationSystem_NNS}), with a tolerance of $10^{-15}$ used in the implementation of Newton's method. Figure~\ref{fig:ErrorPlots1_NNS}b shows the absolute error vs $\eta$ compared with the numerical solution when $\kappa$, $C$, and $E$ are predicted algorithmically for $\alpha=0.8$. The solid curve in Fig.~\ref{fig:ErrorPlots2_NNS} shows the maximum absolute error vs $N$ compared with the numerical solution when $\kappa$, $C$, and $E$ are predicted using the solution of equations system~(\ref{eq:EquationSystem_NNS}). As seen by inspection, the accuracy of the solution, $f(\eta)$, increases with the number of terms used in the series. It is important to note that we do not explicitly enforce that $f(\eta \rightarrow \infty) \to C$ (for all $N$) at $U=0$ (corresponding to $\eta \rightarrow \infty$) in~(\ref{eq:UtTransformedSeriesSolu_NNS}a), as we do for the expansion in~(\ref{eq:AssumedF(U)}) (refer to~(\ref{eq:A0_NNS})). Consequently, for $N = 100$ and 200, this allows for the curves in Fig.~\ref{fig:ErrorPlots1_NNS}b to ultimately attain lower error values than those shown in Fig.~\ref{fig:ErrorPlots1_NNS}a.

The key issue with using the series solution~(\ref{eq:UtTransformedSeriesSolu_NNS}) is its need to access the value of the asymptotic constant $E$. Although not shown here, we have examined a variety of other permissible $\alpha$ values, and we find that the numerical $E$ ultimately becomes negative for $\alpha < 0.74$, thus invalidating the use of the gauge function~(\ref{eq:UTransformation_NNS}) due to a branch point singularity that arises in the physical $\eta$ domain; in this range of $\alpha$, equations system~(\ref{eq:EquationSystem_NNS}) fails to predict converged values for $\kappa$, $C$, and $E$. That said, the asymptotic form with $E$ being negative is perfectly valid for large enough $\eta$, which is the region in which the asymptotic form itself is valid.
\section{Asymptotically Motivated Approximant} \label{sec:Approximant_NNS}
\begin{figure} [htbp]
\centering
    \centering
    \subfloat{{\includegraphics[width=9.5cm]{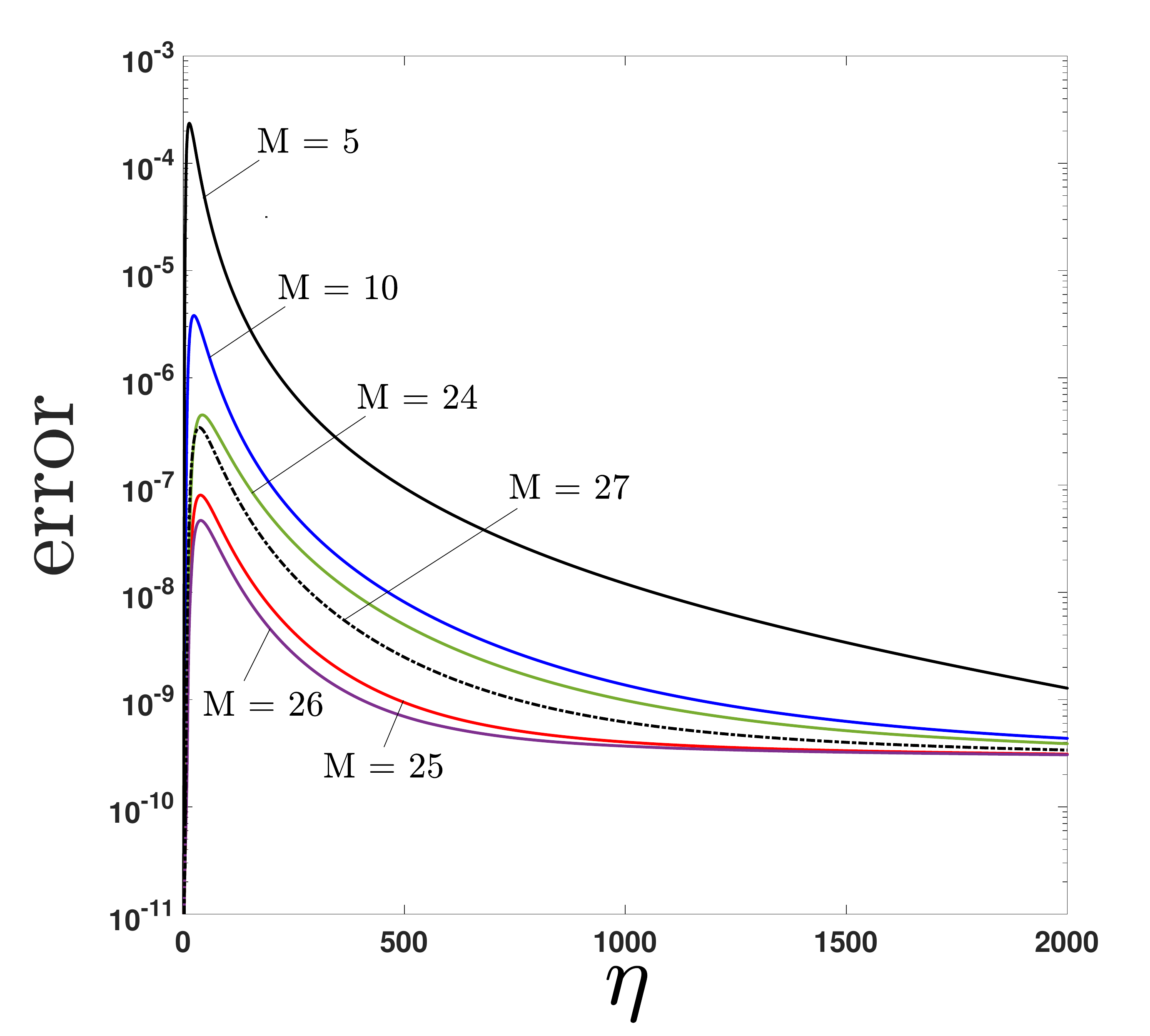} }}
    \caption{Absolute error between $M$-term truncations of the approximant~(\ref{eq:Approximant_NNS}) and the numerical solution (occurring over $\eta \in [0,L]$ with $L=11000$) for $\alpha = 0.8$, plotted versus $\eta$, using $\kappa$ and $C$ generated by the numerical solution. Error decreases until $M=26$, after which it oscillates between curves that are similar to $M = 25$ and 27.}
    \label{fig:ErrorPlot_Approx_pt8}
\end{figure}

\begin{figure} [htbp]
\centering
    \centering
    \subfloat{{\includegraphics[width=9.5cm]{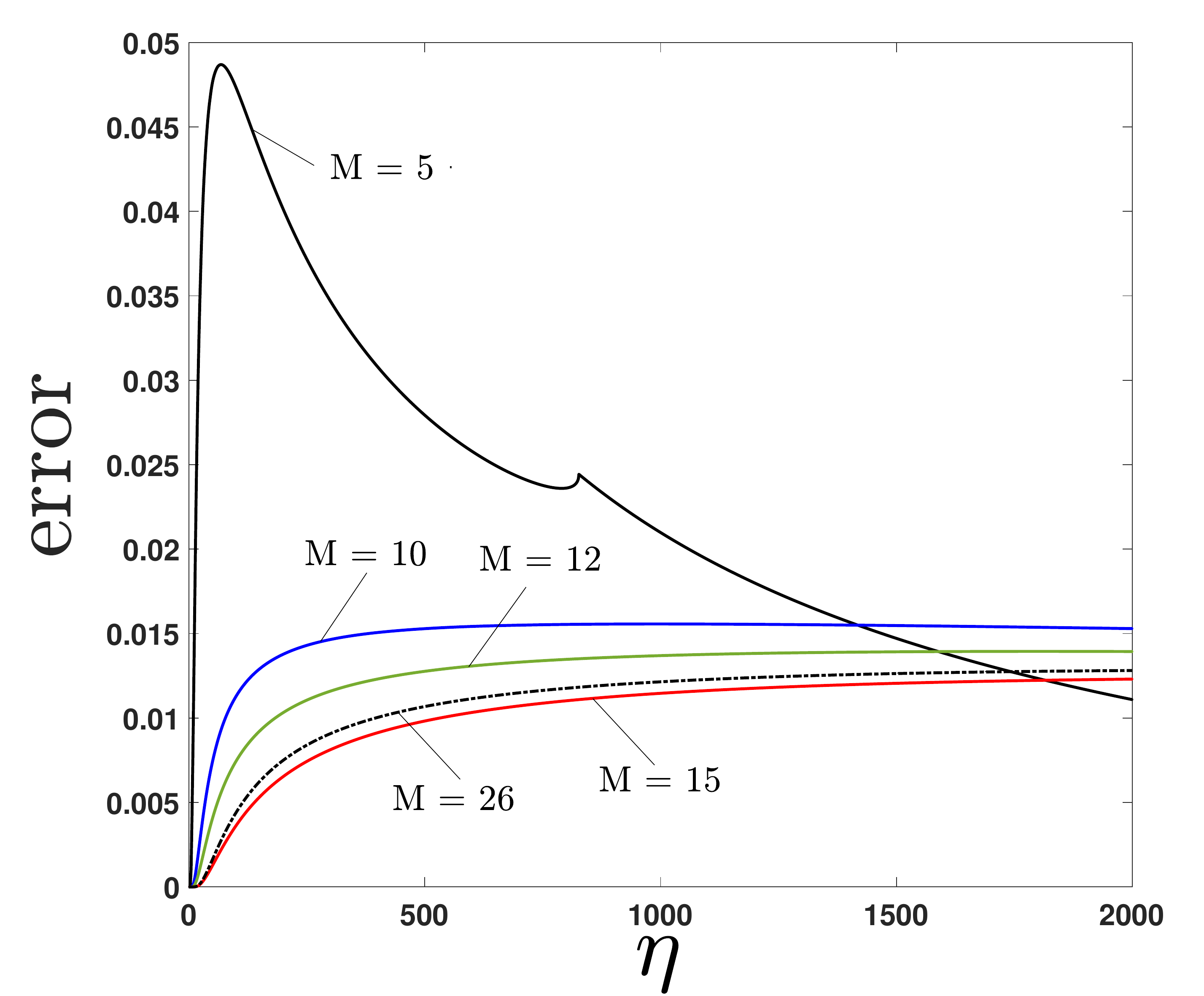} }}
    \caption{Absolute error between $M$-terms truncations of the approximant~(\ref{eq:Approximant_NNS}) and the numerical solution (occurring over $\eta \in [0,L]$ with $L=11000$) for $\alpha = 0.6$, plotted versus $\eta$, using $\kappa$ and $C$ generated by the numerical solution. Error decreases until $M=15$, after which it oscillates between curves that are similar to $M = 12$ and 26.}
    \label{fig:ErrorPlot_Approx_pt6}
\end{figure}

Since a convergent power series solution has only been obtained for $0.74 < \alpha < 1$, we consider an alternative approach to obtain an analytical form over the full range of $0.5 < \alpha < 1$. Note that the form we will obtain can be used as an alternative to the convergent series over the full range of $\alpha$, although some numerical results are needed to do so. One way of overcoming convergence barriers in divergent series solutions is to analytically continue them via Pad\'e approximants~\cite{PadeDefective}. To this end, we utilize an asymptotically motivated approximant\cite{Barlow:2017} in the form of a modified Pad\'e approximant as
\begin{subequations}
\begin{equation}
    f_A = C - \left[ \frac{\displaystyle \sum_{n=0}^{M+1}P_n \eta^n}{\displaystyle \sum_{n=0}^{M}Q_n \eta^n}\right]^\lambda.
\end{equation}
In~(\ref{eq:Approximant_NNS}a), $C$ and $\lambda$ are given by~(\ref{eq:AsymsptoticC_NNS}) and~(\ref{eq:a_and_lambda_NNS}), respectively. To solve for the coefficients $P_n$ and $Q_n$ in~(\ref{eq:Approximant_NNS}a), we write~(\ref{eq:Approximant_NNS}a) in the form
\begin{equation}
    \left[C-\sum_{n=0}^\infty a_n \eta^n \right]^{1/\lambda} = ~ \frac{\displaystyle \sum_{n=0}^{M+1}P_n \eta^n}{\displaystyle \sum_{n=0}^{M}Q_n \eta^n},
\end{equation}
\label{eq:Approximant_NNS}
\end{subequations}
where the coefficients $a_n$ are given in~(\ref{eq:DivSeriesSolu_NNS}a). Using JCP Miller's formula (see Appendix~\ref{sec:JCP}) on the left hand side of~(\ref{eq:Approximant_NNS}b), a standard Pad\'e solver may be employed to solve for the coefficients $P_n$ and $Q_n$ on the right-hand side of~(\ref{eq:Approximant_NNS}b). Note that for $\eta \rightarrow \infty$, $f_A  \sim C - (P_{M+1}/Q_M)^\lambda\eta^\lambda$, which is consistent with the asymptotic form~(\ref{eq:AsymptoticSoluFinal_NNS}) as $\eta \gg E$. To implement the approximant, we use the numerically predicted values of $\kappa$ and $C$, reported in Table~\ref{tab:Table_NumValues}. This is precisely the approach taken by Belden et al.~\cite{FS2020} in the asymptotic approximant solution to the Falkner Skan equation. More details about the numerical prediction for various domain lengths $L$ for $\alpha = 0.8$ are provided in Appendix~\ref{sec:NumericalValues_Table}. As seen in Table~\ref{tab:Table_NumValues}, the numerical solution of the boundary value problem reveals a high sensitivity of parameter values to domain length, and this sensitivity increases as $\alpha$ decreases. Figures~\ref{fig:ErrorPlot_Approx_pt8} and~\ref{fig:ErrorPlot_Approx_pt6} show the absolute error between $M$-term truncations of approximant~(\ref{eq:Approximant_NNS}) and the numerical solution for $\alpha = 0.8$ and $\alpha=0.6$, respectively.  

\begin{table}[htbp]
\begin{center}
\caption{Numerical values of the constants $\kappa$ and $C$ for the non-Newtonian Sakiadis problem. The values are computed using the shooting method algorithm explained in Appendix~\ref{sec:ShootingMethod} with $\eta = \infty$ replaced with a finite surrogate $L$ given in the table. The values of $\kappa$ and $C$ are accurate to within the decimal places reported here, based on convergence by successively increasing $L$.}
\label{tab:Table_NumValues}
\begin{tabular}{ |c|c|l|l| } 
 \hline
 $\alpha$ & $L$ & $~~~~~~~~~~~~~~~~~~\kappa$ & $~~~~~~~~~~~~~~~~~C$ \\ [0.5ex] 
 \hline\hline
 0.99 & 100 & ~-0.4434518189261262~ & ~1.6250769221853265~  \\ 
 \hline
 0.9  & 2000 & ~-0.4413601253597191 & ~1.717915813011322  \\ 
 \hline
 0.8  & 11000 & ~-0.440672715940425 & ~1.860152537  \\
 \hline
 0.7  & 11000& ~-0.442664523 & ~2.08739  \\
 \hline
 0.6  & 11000 & ~-0.44906693 & ~2.56  \\ 
 \hline
~0.55~  & ~40000~ & ~-0.454851 & ~3.1 \\ [1ex] 
 \hline
\end{tabular}
\end{center}
\end{table}

\begin{figure} [htbp]
\centering
    \centering
    \subfloat{{\includegraphics[width=9.5cm]{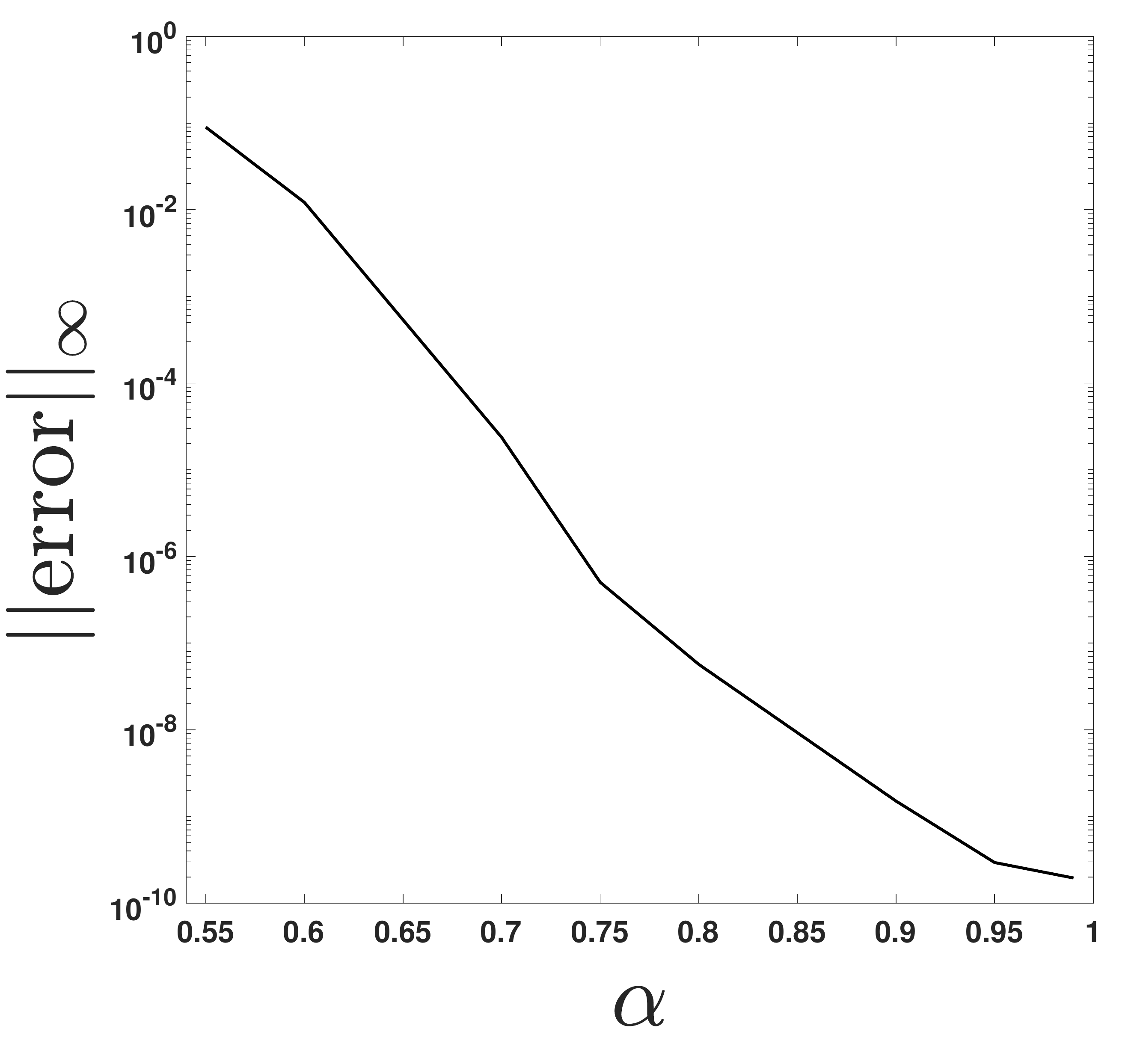} }}
    \caption{Maximum absolute error between approximant~(\ref{eq:Approximant_NNS}) (with $M=20$) and the numerical solution (occurring over $\eta \in [0,L]$) plotted versus $\alpha$, using $\kappa$ and $C$ generated by the numerical solution. The values of $C$ and $\kappa$ are shown in Table~\ref{tab:Table_NumValues}.}
    \label{fig:InfNorm_vs_alpha}
\end{figure}

There are a few defective approximants \footnote{It is possible that poles of a Pad\'e approximant arise within the physical domain of a problem for particular degrees of denominator and numerator. If the exact solution is expected to be finite within the physical domain, these Pad\'es are deemed \textit{defective}. For this exact reason, we did not use $M=25$ in Fig.~\ref{fig:ErrorPlot_Approx_pt6} as it led to a defective approximant for the case of $\alpha = 0.6$.} that arise between the indicated truncations of the approximant in Figs.~\ref{fig:ErrorPlot_Approx_pt8} and~\ref{fig:ErrorPlot_Approx_pt6}, in which the denominator in~(\ref{eq:Approximant_NNS}b) becomes zero for positive $\eta$ values; it is standard practice to ignore these when assessing the solution~\cite{PadeDefective}. We note that the smallest error is obtained at $M=26$ and $M=15$ for the cases of $\alpha = 0.8$ and $0.6$, respectively. For any larger value of $M$, the error oscillates between curves that are similar to $M=25$ and $27$ in Fig.~\ref{fig:ErrorPlot_Approx_pt8} and $M=12$ and $26$ in Fig.~\ref{fig:ErrorPlot_Approx_pt6}. In both cases, we accept the solution as converged, as its precision (defined here as amplitude of these oscillations) is consistent with that of the inputs, particularly $C$ (see Table~\ref{tab:Table_NumValues}). Figure~\ref{fig:InfNorm_vs_alpha} shows the infinity norm (maximum absolute error) vs $\alpha$ for the permissible range of $\alpha$ values when $M = 20$, using the numerical values of the constants $\kappa$ and $C$. 

\section{Post-Processing: Analytically obtained streamlines} \label{sec:StreamlinesPlots_NNS}
\begin{figure} [H]
\centering
    \centering
    \subfloat{{\includegraphics[width=18cm]{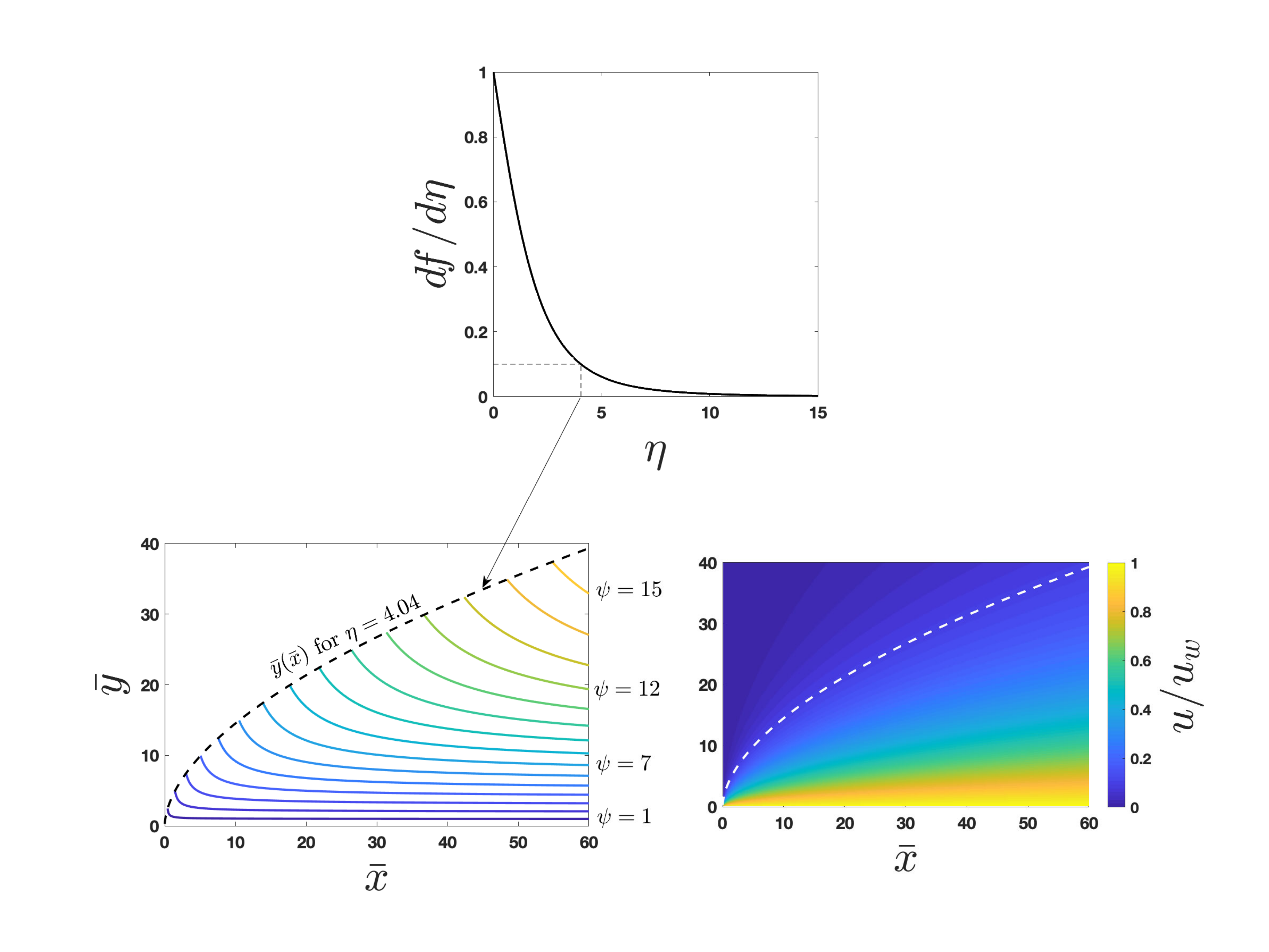} }}
    \caption{ (top) $df/d\eta$ obtained analytically from~(\ref{eq:UTransformedSeriesSolu_NNS}) for $\alpha=0.8$ (using $C$, $\kappa$, and $E$ values from Table~\ref{tab:Table_PredictedValues_Alpha0.8} with $N=400$, using system of equations~(\ref{eq:EquationSystem_NNS})), with gridlines indicating that $df/d\eta=u/u_w=0.1$ at $\eta\approx4.04$. (left) Contours of constant $\psi$ obtained analytically from~(\ref{eq:UTransformedSeriesSolu_NNS}) (using 27 terms) and~(\ref{eq:psi}), displayed in increments of $\Delta\psi=1$ in the $\bar{y}\equiv u_w y$ vs. $\bar{x}\equiv K u_w^{2\alpha-1} x/\rho$ plane. The dashed curve is the envelope of the boundary layer (chosen here to be the locus of points at which $u/u_w = 0.1$), and restricts the display of the streamlines for velocities where $u/u_w > 0.1$. (right) The $u$ velocity field obtained analytically from~(\ref{eq:UTransformedSeriesSolu_NNS}) (using 27 terms) and~(\ref{eq:u}).}
    \label{fig:streamlines}
\end{figure}

Now that we have accurate analytical solutions to~(\ref{eq:ODESystem_NNS}), we may insert $f$, given by either~(\ref{eq:UTransformedSeriesSolu_NNS}) (for $0.74\le\alpha \leq1$)\footnote{For $\alpha=1$, one can use the Newtonian result~(\ref{eq:Solution_NS}) in place of (\ref{eq:UTransformedSeriesSolu_NNS}); coefficients for~(\ref{eq:Solution_NS}) are provided by Naghshineh et al\cite{FlatWallSakiadisPaper}.} or~(\ref{eq:Approximant_NNS}) (for $0.5<\alpha<1$), and its derivative $f'(\eta)$ (which may obtained analytically) into~(\ref{eq:u})  to obtain the $u$ velocity field, which is shown in the right-hand plot in Fig.~\ref{fig:streamlines}. The paths of fluid points, i.e., the streamlines of constant $\psi$, may be extracted easily from the analytical solution. To do so, we explicitly solve for the $x$ and $y$ coordinates of a given streamline $\psi =$~constant by rearranging the equations in~(\ref{eq:psi&eta}) to yield:
\begin{equation}
y=\frac{\psi\eta}{u_w f(\eta)},~~x=\frac{\rho}{Ku_w^{2\alpha-1}}\left(\frac{\psi}{f(\eta)}\right)^{\alpha+1}.
\label{eq:xandy}
\end{equation}
Equation~(\ref{eq:xandy}) provides a parametric representation of the streamlines in terms of $\eta$ and $f(\eta)$, the latter given analytically by~(\ref{eq:UTransformedSeriesSolu_NNS}) or~(\ref{eq:Approximant_NNS}).  Figure~\ref{fig:streamlines} provides a typical streamline plot extracted in this way.  In the figure, a dashed curve plots the boundary layer thickness $y=\eta\left(K~x~u_w^{\alpha-2}/\rho\right)^{1/(\alpha+1}$, defined here as the locus of points where the fluid velocity is reduced to 10\% of the wall velocity; according to~(\ref{eq:eta}), this occurs when $u/u_w=df/d\eta=0.1$. From the topmost plot of Fig.~\ref{fig:streamlines}, this occurs when $\eta=4.04$. The benefit of the analytical solution is clearly indicated here, as streamline plots can be generated accurately to any desired resolution with low computational cost.

\section{Conclusions} \label{sec:Conclusion_NNP}
In this work, we provide a convergent power series solution to the non-Newtonian Sakiadis boundary layer problem, valid for $0.74 \leq \alpha < 1$, using the asymptotic expansion as $\eta \to \infty$ to determine a gauge function for the series. The asymptotically motivated series fails when the gauge function is unable to completely transverse the physical domain due to a branch point singularity that arises in the asymptotic form. We note that although we developed an asymptotic approximant to model cases where $0.5 < \alpha < 0.74$, the approximant is capable of representing the solution for all $\alpha$ values in the range $0.5 < \alpha < 1$. Once obtained, the analytical solutions enable computationally-efficient post-processing to extract streamlines to any desired resolution.

\appendix
\section{Useful Formulae for Manipulating Series}
\label{sec:JCP&Cauchy}
\subsection{Raising a series to a power \label{sec:JCP}}
The following relation is JCP Miller's formula for raising a series to a power~\cite{Henrici}:
\begin{subequations}
\begin{equation}
    \left(\sum_{n=0}^{\infty}a_{n}x^{n}\right)^\gamma = \sum_{n=0}^{\infty}b_{n}x^{n}, 
\end{equation}
where
\begin{equation}
   b_{n>0} = \frac{1}{n~a_{0}}\sum_{j=1}^{n}(j\gamma-n+j) a_j b_{n-j}~,~b_0 = \left(a_0\right)^\gamma~,~a_0 \neq 0. 
\end{equation}
\end{subequations}
\subsection{Product of two series \label{sec:Cauchy}}
The following relation is the well-known Cauchy product of two series~\cite{Churchill}:
\begin{equation}
    \sum_{n=0}^{\infty}a_{n}x^{n}\sum_{n=0}^{\infty}b_{n}x^{n} = \sum_{n=0}^{\infty}\left(\sum_{j=0}^{n}a_{j}b_{n-j}\right)x^{n}.
    \label{eq:Cauchy}
\end{equation}
\section{\quad Numerical Solution: Shooting Method \label{sec:ShootingMethod}}
The algorithm below is motivated from the work Cebeci et al.~\cite{Cebeci}, developed for the Falkner-Skan boundary layer problem. Here, we extend their approach to the non-Newtonian Sakiadis problem given by~(\ref{eq:ODESystem_NNS}).  With the goal of determining the value of $\kappa$ defined by~(\ref{eq:kappa}), we approximate the boundary value problem~(\ref{eq:ODESystem_NNS}) in $f(\eta)$ (defined on a semi-infinite domain) with the following initial value problem (IVP) in $f(\eta;\kappa)$ (defined on a finite domain):
\begin{subequations}
\begin{equation}
    \alpha (\alpha+1)f''' - (-f'')^{(2-\alpha)}f = 0,~~0\le\eta\le L
    \label{eq:ODE_shooting}
\end{equation}
with the initial conditions (taken from~(\ref{eq:f(0)BC_SP_NNS}), (\ref{eq:f'(0)BC_SP_NNS}), and~(\ref{eq:kappa}))
\begin{equation}
    f(0;\kappa) = 0~~,~~f'(0;\kappa) = 1~~,~~f''(0;\kappa) = \kappa,
    \label{eq:BCs_shooting}
\end{equation}
\label{eq:finiteODE}
\end{subequations}
where $f'$ denotes the derivative of $f$ with respect to $\eta$. In order to determine $\kappa$, we subject the IVP~(\ref{eq:finiteODE}) to the constraint
\begin{equation}
f'(L;\kappa)=0,
\label{eq:constraint}
\end{equation}
which incorporates condition~(\ref{eq:f'(inf)BC_SP_NNS}) such that the solution to~(\ref{eq:finiteODE}) limits to the solution of~(\ref{eq:ODESystem_NNS}) as $L\to\infty$.  The determination of $\kappa$ and the numerical solution for $f$ itself from~(\ref{eq:finiteODE}) (subject to~(\ref{eq:constraint})) is obtained by the method of shooting, as outlined below. 

First, we replace~(\ref{eq:ODE_shooting}) with a system of three first-order ODEs. To do so, we let $\mathcal{U}(\eta;\kappa)$, $\mathcal{V}(\eta;\kappa)$, and $\mathcal{V}'(\eta;\kappa)$ represent $f'(\eta;\kappa)$, $f''(\eta;\kappa)$, and $f'''(\eta;\kappa)$, respectively. Thus,~(\ref{eq:finiteODE}) can be written as the system
\begin{subequations}
\begin{equation}
   f' = \mathcal{U}~~,~~\mathcal{U}' = \mathcal{V}~~,~~\mathcal{V}' = \frac {(-\mathcal{V})^{(2-\alpha)}f} {\alpha(\alpha+1)},
\label{eq:First3equa_shooting}
\end{equation}
with initial conditions
\begin{equation}
    f(0;\kappa) = 0~~,~~\mathcal{U}(0;\kappa) = 1~~,~~ \mathcal{V}(0;\kappa) = \kappa.
\label{eq:IVP1_conditions}
\end{equation}
\label{eq:IVP_set1_shooting}
\end{subequations}

The objective is to provide a solution to the IVP~(\ref{eq:IVP_set1_shooting}), such that the constraint (\ref{eq:constraint}) is satisfied. That is, we solve for the solution of~(\ref{eq:IVP_set1_shooting}) by seeking $\kappa$, such that
\begin{equation}
    \mathcal{U}(L; \kappa)  = 0,
\end{equation}
where $L$ is successively increased, such that the $\kappa$ value for~(\ref{eq:finiteODE}) approaches the $\kappa$ value for~(\ref{eq:ODESystem_NNS}).

In order to determine $\kappa$ in~(\ref{eq:IVP1_conditions}), we use Newton's method~\cite{Newton'sMethod} defined by
\begin{equation}
\nonumber
    \kappa^{\gamma+1} = \kappa^{\gamma} - \frac{\mathcal{U}(L; \kappa^\gamma)}{\frac{\partial }{\partial \kappa} (\mathcal{U}(L;\kappa^\gamma))}~~,~~ \gamma = 0,~1,~2,~...,
\end{equation}
with $\kappa^0$ being the initial estimate for $\kappa$, and $\gamma$ is the iteration number. In order to obtain the derivative of $\mathcal{U}$ with respect to $\kappa$, we take the derivative of~(\ref{eq:IVP_set1_shooting}) with respect to $\kappa$, which leads to the following additional IVP:
\begin{subequations}
\begin{equation}
   \frac{\partial f'}{\partial\kappa} = \frac{\partial\mathcal{U}}{\partial \kappa}~~,~~\frac{\partial \mathcal{U}'}{\partial\kappa} = \frac{\partial\mathcal{V}}{\partial\kappa}~~,~~\frac{\partial\mathcal{V}'}{\partial\kappa} = (\alpha-2)\frac{\partial\mathcal{V}}{\partial\kappa}~\frac{(-\mathcal{V})^{(1-\alpha)}}{\alpha(\alpha+1)}f +\frac{(-\mathcal{V})^{(2-\alpha)}}{\alpha(\alpha+1)}~\frac{\partial f}{\partial\kappa},
    \label{eq:IVP_set2_shooting}
\end{equation}   
with conditions
\begin{equation}
   \frac{\partial f}{\partial\kappa}(0;\kappa) = 0~~,~~ \frac{\partial \mathcal{U}}{\partial\kappa}(0;\kappa) = 0~~,~~\frac{\partial \mathcal{V}}{\partial\kappa}(0;\kappa) = 1.
    \label{eq:IVP_BCs_set2_Shooting}
\end{equation}
\label{eq:IVP_set2_shooting}
\end{subequations}

For clarity, we assign new variables to the derivatives with respect to $\kappa$ as follows
\[\mathfrak{F}(\eta;\kappa)\equiv\frac{\partial f}{\partial\kappa}(\eta;\kappa),~~\mathfrak{U}(\eta;\kappa)\equiv\frac{\partial \mathcal{U}}{\partial\kappa}(\eta;\kappa),~~\mathfrak{V}(\eta;\kappa)\equiv\frac{\partial \mathcal{V}}{\partial\kappa}(\eta;\kappa),\]
 commute the differentiation with respect to $\eta$ (denoted by primes) and differentiation with respect to $\kappa$ in~(\ref{eq:IVP_set2_shooting}), and combine~(\ref{eq:IVP_set1_shooting}) and~(\ref{eq:IVP_set2_shooting}) into the single IVP evaluated at $\kappa=\kappa^\gamma$:
 \begin{equation}
\frac{d}{d\eta}\left[ \begin{array}{ccccccc}
    f\\ \mathcal{U} \\ \mathcal{V}\\ \mathfrak{F} \\ \mathfrak{U} \\ \mathfrak{V} \end{array} \right]=\left[ \begin{array}{ccccccc}   
\mathcal{U} \\ \mathcal{V}\\ \frac {(-\mathcal{V})^{(2-\alpha)}f} {\alpha(\alpha+1)}\\ \mathfrak{U} \\ \mathfrak{V}\\ (\alpha-2)\mathfrak{V}\frac{(-\mathcal{V})^{(1-\alpha)}}{\alpha(\alpha+1)}f +\frac{(-\mathcal{V})^{(2-\alpha)}}{\alpha(\alpha+1)}\mathfrak{F}
    \end{array} \right],~~\left[ \begin{array}{ccccccc}
    f\\ \mathcal{U} \\ \mathcal{V}\\ \mathfrak{F} \\ \mathfrak{U} \\ \mathfrak{V} \end{array} \right]_{\eta=0}=\left[ \begin{array}{ccccccc}
    0\\ 1 \\ \kappa^\gamma\\ 0 \\ 0 \\ 1 \end{array} \right].
    \label{eq:fullsystem}
 \end{equation}
From~(\ref{eq:fullsystem}), both $\mathcal{U}(L; \kappa^\gamma)$ and $\mathfrak{U}(L; \kappa^\gamma)$ may be determined; then the guess $\kappa^\gamma$ may be progressed to the next iteration via
\begin{equation}
    \kappa^{\gamma+1} = \kappa^{\gamma} - \frac{\mathcal{U}(L; \kappa^\gamma)}{\mathfrak{U}(L;\kappa^\gamma)}~~,~~ \gamma = 0,~1,~2,~...,
    \label{eq:NewtonsMethod_shooting}
\end{equation}
to compute $\kappa ^ {\gamma + 1}$. We use a fourth-order Runge-Kutta method, with $\Delta\eta=0.001$ to solve~(\ref{eq:fullsystem}). A convergence requirement of $|\kappa^{\gamma+1}-\kappa^\gamma|<10^{-15}$ is enforced in the Newton iteration~(\ref{eq:NewtonsMethod_shooting}).
\section{\quad Evaluations of $\kappa$, $C$, and $E$ \label{sec:AsymptoticConstants_Table}}
\subsection{Numerical evaluation of $E$ \label{sec:E_numerical}}
In this section, we show that the constant $E \to 1$ as $\alpha \to 1$. These values are obtained numerically via a shooting method described in Appendix~\ref{sec:ShootingMethod} combined with the definition of $E$ defined in~(\ref{eq:E_NNS}).

\begin{table}[htbp]
\begin{center}
\caption{Numerical values of the constant $E$ defined in~(\ref{eq:E_NNS}). The values are computed using the shooting method algorithm explained in Appendix~\ref{sec:ShootingMethod} with $\eta = \infty$ replaced with a finite surrogate $L$ given in the table. }
\label{tab:Table_NumE}
\begin{tabular}{ |c|c|l| } 
 \hline
 $\alpha$ & $L$ & $~~~~~~~~~~~~~~~~E$ \\ [0.5ex] 
 \hline\hline
 ~0.99~ & ~100~ & ~0.996357144778925~~\\ 
 \hline
 0.9 & ~2000~ & ~0.921210805240775~~\\
 \hline
 0.8 & ~11000~ & ~0.639133894794667~~\\
 \hline
 0.75 & ~11000~ & ~0.249305949228073~~\\ 
 [1ex] 
 \hline
\end{tabular}
\end{center}
\end{table}

\subsection{\quad Evaluations of $\kappa$, $C$, and $E$ for $\alpha = 0.8$\label{sec:NumericalValues_Table}}
In what follows, we provide constants used in the analytical solutions~(\ref{eq:DivSeriesSolu_NNS}),~(\ref{eq:UTransformation_NNS}),~(\ref{eq:UtTransformedSeriesSolu_NNS}), and~(\ref{eq:Approximant_NNS}) for $\alpha = 0.8$. These are obtained in two ways--either numerically via a shooting method (Table~\ref{tab:Table_NumValues_Alpha0.8} and Appendix~\ref{sec:ShootingMethod}), or via a self-contained algorithm using the power series (Table~\ref{tab:Table_PredictedValues_Alpha0.8} and Appendix~(\ref{sec:PredictionAlgorithm_Appendix})). 

\begin{table}[htbp]
\begin{center}
\caption{Numerical values of the constants $\kappa$, $C$, and $E$ with $\alpha = 0.8$ for the non-Newtonian Sakiadis problem. The values are computed using the shooting method algorithm explained in Appendix~\ref{sec:ShootingMethod} with $\eta = \infty$ replaced with a finite surrogate $L$ given in the table. This numerical algorithm uses a Newton iteration with a tolerance of $10^{-15}$, wrapped around a 4$^\mathrm{th}$-order Runge-Kutta solver with a step size of $\triangle \eta = 0.001$. }
\label{tab:Table_NumValues_Alpha0.8}
\begin{tabular}{ |c|l|l|l| } 
 \hline
 $L$ & $~~~~~~~~~~~~~~~\kappa$ & $~~~~~~~~~~~~~~~C$ & $~~~~~~~~~~~~~~~~E$ \\ [0.5ex] 
 \hline\hline
 40 & ~-0.44069277587148~ & ~1.85680030949222~ & ~0.654388901675618~\\ 
 \hline
 100 & ~-0.44067330624905~ & ~1.85989634054125~ & ~0.641907516688146~\\
 \hline
 240 & ~-0.44067273476661~ & ~1.86013263767144~ & ~0.639639542413533~\\
 \hline
 540 & ~-0.44067271669187~ & ~1.86015073982904~ & ~0.639235534657387~\\ 
 \hline
1000 & ~-0.44067271600486~ & ~1.86015225132246~ & ~0.639163575176951~\\ 
 \hline
2000 & ~-0.44067271594447~  & ~1.86015250138706~ & ~0.639141167249476~ \\ 
 \hline
5000  & ~-0.44067271594052~ & ~1.86015253504567~ & ~0.639134860809690~ \\ 
 \hline
~11000~ & ~-0.44067271594043~ & ~1.86015253716140~ & ~0.639133894794667~  \\ [1ex] 
 \hline
\end{tabular}
\end{center}
\end{table}

\begin{table}[htbp]
\begin{center}
\caption{Predicted values of the constants $\kappa$, $C$, and $E$ with $\alpha = 0.8$ for the non-Newtonian Sakiadis problem. The values are computed using the the solution of equations system~(\ref{eq:EquationSystem_NNS}) explained in Sec.(\ref{sec:PredictionAlgorithm}) with the $N$ values given in the table. This algorithm uses a Newton iteration with a tolerance of $10^{-15}$.}
\label{tab:Table_PredictedValues_Alpha0.8}
\begin{tabular}{ |c|l|l|l| } 
 \hline
 $N$ & $~~~~~~~~~~~~~~~~\kappa$ & $~~~~~~~~~~~~~~~~C$ & $~~~~~~~~~~~~~~~~E$ \\ [0.5ex] 
 \hline\hline
 50 & ~-0.440672770686287~ & ~1.860151847487335~ & ~0.639137961392650~\\
 \hline
 100 & ~-0.440672715934279  & ~1.860152537361902 & ~0.639133656731447\\
 \hline
 200 & ~-0.440672715934271 & ~1.860152537362062 & ~0.639133656729539\\
 \hline
 ~400~ & ~-0.440672715934270~ & ~1.860152537362068~ & ~0.639133656729520~~\\ [1ex] 
 \hline
\end{tabular}
\end{center}
\end{table}
\section{\quad Newton's Method Used to Predict Constants
\label{sec:PredictionAlgorithm_Appendix}}
The system of equations used in this algorithm are:
\begin{subequations}
    \begin{equation}
        \mathcal{F}_1(\kappa, C, E) = \left[ \sum_{n=0}^N \hat{A}_n (-1)^n\right]-C,
        \label{eq:F1_Appendix}
    \end{equation}
    \begin{equation}
        \mathcal{F}_2(\kappa, C, E) = \left[ \sum_{n=0}^N n\hat{A}_n (-1)^{n-1}\right]-A_1,
        \label{eq:F2_Appendix}
    \end{equation}
    \begin{equation}
        \mathcal{F}_3(\kappa, C, E) = \hat{A}_N,
        \label{eq:F3_Appendix}
\end{equation}  
\end{subequations}
where $C$, $A_1$, and $\hat{A}_N$ are defined in~(\ref{eq:AsymsptoticC_NNS}), (\ref{eq:A1_NNS}), and~(\ref{eq:UtTransformedSeriesSolu_NNS}g), respectively.

We take the derivatives of~(\ref{eq:F1_Appendix})-(\ref{eq:F3_Appendix}) with respect to $\kappa$, $C$, and $E$, and obtain the Jacobian matrix,
\begin{equation}
    \textbf{J} = \left[ \begin{array}{ccccccc}
    \frac{\partial \mathcal{F}_1}{\partial C} &&& \frac{\partial \mathcal{F}_1}{\partial E} &&& \frac{\partial \mathcal{F}_1}{\partial \kappa}  \\ 
    \frac{\partial \mathcal{F}_2}{\partial C} &&& \frac{\partial \mathcal{F}_2}{\partial E} &&& \frac{\partial \mathcal{F}_2}{\partial \kappa} \\ 
    \frac{\partial \mathcal{F}_3}{\partial C} &&& \frac{\partial \mathcal{F}_3}{\partial E} &&& \frac{\partial \mathcal{F}_3}{\partial \kappa} \end{array} \right],
    \label{eq:Jacobian}
\end{equation}
which is used in Newton’s Method
\begin{equation}
    \left[ \begin{array}{ccccccc}
    C^{j+1} \\ E^{j+1} \\  \kappa^{j+1} \end{array} \right]  =  \left[ \begin{array}{ccccccc} 
    C^{j} \\ E^{j} \\  \kappa^{j} \end{array} \right] - \textbf{J}^{-1} 
    \left[ \begin{array}{ccccccc}
    \mathcal{F}_1(C^{j}, E^{j},\kappa^{j}) \\
    \mathcal{F}_2(C^{j}, E^{j},\kappa^{j}) \\
    \mathcal{F}_3(C^{j}, E^{j},\kappa^{j}) \\
    \end{array} \right].
    \label{eq:NewtonsMethod_NNS}
\end{equation}
To construct the Jacobian matrix~(\ref{eq:Jacobian}), we take the derivative of~(\ref{eq:F1_Appendix} -~(\ref{eq:F3_Appendix}), with respect to $C$, $E$, and $\kappa$, as follows:
\begin{subequations}
\begin{equation}
    \frac{\partial \mathcal{F}_1}{\partial C}=\left[\sum_{n=0}^\infty \frac{\partial \hat{A}_n}{\partial C}(-1)^n\right]-1~~,~~\frac{\partial \mathcal{F}_1}{\partial E}=\sum_{n=0}^\infty \frac{\partial \hat{A}_n}{\partial E}(-1)^n~~,~~\frac{\partial \mathcal{F}_1}{\partial \kappa}=\sum_{n=0}^\infty \frac{\partial \hat{A}_n}{\partial \kappa}(-1)^n,
\end{equation}
\begin{equation}
    \frac{\partial \mathcal{F}_2}{\partial C}=\left[\sum_{n=0}^\infty n\frac{\partial \hat{A}_n}{\partial C}(-1)^{n-1}\right]-\frac{\partial A_1}{\partial C},
\end{equation}
\begin{equation}
    \frac{\partial \mathcal{F}_2}{\partial E}=\left[\sum_{n=0}^\infty n\frac{\partial \hat{A}_n}{\partial E}(-1)^{n-1}\right]-\frac{\partial A_1}{\partial E},
\end{equation}
\begin{equation}
    \frac{\partial \mathcal{F}_2}{\partial \kappa}=\left[\sum_{n=0}^\infty n\frac{\partial \hat{A}_n}{\partial \kappa}(-1)^{n-1}\right]-\frac{\partial A_1}{\partial \kappa},
\end{equation}
\begin{equation}
    \frac{\partial \mathcal{F}_3}{\partial C}=\frac{\partial \hat{A}_N}{\partial C}~~,~~\frac{\partial \mathcal{F}_3}{\partial E}=\frac{\partial \hat{A}_N}{\partial E}~~,~~\frac{\partial \mathcal{F}_3}{\partial \kappa}=\frac{\partial \hat{A}_N}{\partial \kappa}.
\end{equation}
\end{subequations}

We take the derivative of~(\ref{eq:a_and_lambda_NNS}) with respect to the unknowns $C$, $E$, and $\kappa$. It is important to note that $\lambda$ is not function of any these three constants. Hence, we only take the derivative of $\mathcal{A}$ in~(\ref{eq:a_and_lambda_NNS}) as follows:
\begin{equation}
    \frac{\partial \mathcal{A}}{\partial C} = \frac{1-\alpha}{\alpha(\alpha+1)E}~~,~~\frac{\partial \mathcal{A}}{\partial E} = \frac{-C(1-\alpha)}{\alpha(\alpha+1)E^2}~~,~~\frac{\partial \mathcal{A}}{\partial \kappa} = 0.
\end{equation}

We use the chain rule to take the derivatives of~(\ref{eq:k1})-(\ref{eq:k5}), as follows:
\begin{subequations}
\begin{equation}
   \frac{\partial k_1}{\partial C} = 3\alpha(\alpha+1)\lambda^3\mathcal{A}^2\frac{\partial \mathcal{A}}{\partial C}~~,~~\frac{\partial k_1}{\partial E} = 3\alpha(\alpha+1)\lambda^3\mathcal{A}^2\frac{\partial \mathcal{A}}{\partial E}~~,~~\frac{\partial k_1}{\partial \kappa} = 0,
\end{equation}
\begin{equation}
        \frac{\partial k_2}{\partial C} = 9\alpha(\alpha+1)\lambda^2(\lambda-1)\mathcal{A}^2\frac{\partial \mathcal{A}}{\partial C}~~,~~   \frac{\partial k_2}{\partial E} = 9\alpha(\alpha+1)\lambda^2(\lambda-1)\mathcal{A}^2\frac{\partial \mathcal{A}}{\partial E}~~,~~ \frac{\partial k_2}{\partial \kappa} = 0,
\end{equation}
\begin{equation}
    \frac{\partial k_3}{\partial C} = 3\alpha(\alpha+1)\lambda(\lambda-1)(\lambda-2)\mathcal{A}^2\frac{\partial \mathcal{A}}{\partial C},
\end{equation}
\begin{equation}
    \frac{\partial k_3}{\partial E} = 3\alpha(\alpha+1)\lambda(\lambda-1)(\lambda-2)\mathcal{A}^2\frac{\partial \mathcal{A}}{\partial E}~~,~~ \frac{\partial k_3}{\partial \kappa} = 0,
\end{equation}
\begin{equation}
    \frac{\partial k_4}{\partial C} = -2\lambda^2\mathcal{A}\frac{\partial \mathcal{A}}{\partial C}~~,~~\frac{\partial k_4}{\partial E} = -2\lambda^2\mathcal{A}\frac{\partial \mathcal{A}}{\partial E}~~,~~\frac{\partial k_4}{\partial \kappa} = 0,
\end{equation}
\begin{equation}
    \frac{\partial k_5}{\partial C} = -2\lambda(\lambda-1)\mathcal{A}\frac{\partial \mathcal{A}}{\partial C}~~,~~\frac{\partial k_5}{\partial E} = -2\lambda(\lambda-1)\mathcal{A}\frac{\partial \mathcal{A}}{\partial E}~~,~~ \frac{\partial k_5}{\partial \kappa} = 0.
\end{equation}
\end{subequations}

We take derivatives of $\hat{A}_0$ through $\hat{A}_4$ in~(\ref{eq:UtTransformedSeriesSolu_NNS}b)-(\ref{eq:UtTransformedSeriesSolu_NNS}f), as follows
\begin{subequations}
\begin{equation}
        \frac{\partial \hat{A}_0}{\partial C} = 0~~,~~\frac{\partial \hat{A}_0}{\partial E} = 0 ~~,~~\frac{\partial \hat{A}_0}{\partial \kappa} = 0.
\end{equation}
\begin{equation}
        \frac{\partial \hat{A}_1}{\partial C} = \frac{-1}{\lambda \mathcal{A}^2}\frac{\partial \mathcal{A}}{\partial C}~~,~~\frac{\partial \hat{A}_1}{\partial E} = \frac{-1}{\lambda \mathcal{A}^2}\frac{\partial \mathcal{A}}{\partial E}~~,~~\frac{\partial \hat{A}_1}{\partial \kappa} = 0,
\end{equation}
\begin{equation}
        \frac{\partial \hat{A}_2}{\partial C} = \frac{1}{2\lambda^2}\left(   
        \frac{-2\kappa}{a^3}+\frac{\lambda-1}{\mathcal{A}^2}\right)\frac{\partial \mathcal{A}}{\partial C},
\end{equation}
\begin{equation}
        \frac{\partial \hat{A}_2}{\partial E} = \frac{1}{2\lambda^2}\left(   
        \frac{-2\kappa}{a^3}+\frac{\lambda-1}{\mathcal{A}^2}\right)\frac{\partial \mathcal{A}}{\partial E}~~,~~\frac{\partial \hat{A}_2}{\partial \kappa} = \frac{1}{2\lambda^2\mathcal{A}^2}.
\end{equation}
\begin{equation}
    \frac{\partial \hat{A}_3}{\partial C} = u_1\frac{\partial v_1}{\partial C} + v_1\frac{\partial u_1}{\partial C}~~,~~ \frac{\partial \hat{A}_3}{\partial E} = u_1\frac{\partial v_1}{\partial E} + v_1\frac{\partial u_1}{\partial E}~~,~~\frac{\partial \hat{A}_3}{\partial \kappa} = u_1\frac{\partial v_1}{\partial \kappa},
\end{equation}
where
\begin{equation}
    u_1 = \frac{1}{6k_1}~~,~~v_1 = -2k_2\hat{A}_2 - k_3 \hat{A}_1,
\end{equation}
\begin{equation}
    \frac{\partial u_1}{\partial C}=\frac{-1}{6(k_1)^2}\frac{\partial k_1}{\partial C}~~,~~\frac{\partial u_1}{\partial E}=\frac{-1}{6(k_1)^2}\frac{\partial k_1}{\partial E}~~,~~\frac{\partial u_1}{\partial \kappa}=0,
\end{equation}
\begin{equation}
    \frac{\partial v_1}{\partial C}=-2\left(k_2\frac{\partial \hat{A}_2}{\partial C}+\hat{A}_2\frac{\partial k_2}{\partial C}\right)-\left(k_3\frac{\partial \hat{A}_1}{\partial C}+ \hat{A}_1\frac{\partial k_3}{\partial C}\right),
\end{equation}
\begin{equation}
    \frac{\partial v_1}{\partial E}=-2\left(k_2\frac{\partial \hat{A}_2}{\partial E}+\hat{A}_2\frac{\partial k_2}{\partial E}\right)-\left(k_3\frac{\partial \hat{A}_1}{\partial E}+ \hat{A}_1\frac{\partial k_3}{\partial E}\right),
\end{equation}
\begin{equation}
    \frac{\partial v_1}{\partial \kappa}=-2k_2\frac{\partial \hat{A}_2}{\partial \kappa}.
\end{equation}
\begin{equation}
    \frac{\partial \hat{A}_4}{\partial C} = u_2\frac{\partial v_2}{\partial C} + v_2\frac{\partial u_2}{\partial C}~~,~~\frac{\partial \hat{A}_4}{\partial E} = u_2\frac{\partial v_2}{\partial E} + v_2\frac{\partial u_2}{\partial E}~~,~~\frac{\partial \hat{A}_4}{\partial \kappa} = u_2\frac{\partial v_2}{\partial \kappa},
\end{equation}
and
\begin{equation}
    u_2=\frac{1}{24k_1}~~,~~v_2=-2k_2\hat{A}_2 - 6k_2\hat{A}_3-2k_3\hat{A}_2+\hat{A}_1\hat{d}_0-12k_1\hat{A}_3,
\end{equation}
\begin{equation}
    \frac{\partial u_2}{\partial C}=\frac{-1}{24(k_1)^2}\frac{\partial k_1}{\partial C}~~,~~\frac{\partial u_2}{\partial E}=\frac{-1}{24(k_1)^2}\frac{\partial k_1}{\partial E}~~,~~\frac{\partial u_2}{\partial \kappa}=0,
\end{equation}
\begin{multline}
     \frac{\partial v_2}{\partial C}=-2\left(k_2\frac{\partial \hat{A}_2}{\partial C}+\hat{A}_2\frac{\partial k_2}{\partial C}\right)-6\left(k_2\frac{\partial \hat{A}_3}{\partial C}+\hat{A}_3\frac{\partial k_2}{\partial C}\right)-2\left(k_3\frac{\partial \hat{A}_2}{\partial C}+\hat{A}_2\frac{\partial k_3}{\partial C}\right)+\\
     \hat{A}_1\frac{\partial \hat{d}_0}{\partial C}+\hat{d}_0\frac{\partial \hat{A}_1}{\partial C}-12\left(k_1\frac{\partial \hat{A}_3}{\partial C}+\hat{A}_3\frac{\partial k_1}{\partial C}\right),
\end{multline}
\begin{multline}
     \frac{\partial v_2}{\partial E}=-2\left(k_2\frac{\partial \hat{A}_2}{\partial E}+\hat{A}_2\frac{\partial k_2}{\partial E}\right)-6\left(k_2\frac{\partial \hat{A}_3}{\partial E}+\hat{A}_3\frac{\partial k_2}{\partial E}\right)-2\left(k_3\frac{\partial \hat{A}_2}{\partial E}+\hat{A}_2\frac{\partial k_3}{\partial E}\right)+\\
     \hat{A}_1\frac{\partial \hat{d}_0}{\partial E}+\hat{d}_0\frac{\partial \hat{A}_1}{\partial E}-12\left(k_1\frac{\partial \hat{A}_3}{\partial E}+\hat{A}_3\frac{\partial k_1}{\partial E}\right),
\end{multline}
\begin{equation}
    \frac{\partial v_2}{\partial \kappa}=-2k_2\frac{\partial \hat{A}_2}{\partial \kappa}-6k_2\frac{\partial \hat{A}_3}{\partial \kappa}-2k_3\frac{\partial \hat{A}_2}{\partial \kappa}+\hat{A}_1\frac{\partial \hat{d}_0}{\partial \kappa}-12k_1\frac{\partial \hat{A}_3}{\partial \kappa},
\end{equation}
\begin{equation}
    \frac{\partial \hat{d}_0}{\partial C}=(2-\alpha)(\hat{c}_0)^{1-\alpha}\frac{\partial \hat{c}_0}{\partial C},
\end{equation}
\begin{equation}
    \frac{\partial \hat{d}_0}{\partial E}=(2-\alpha)(\hat{c}_0)^{1-\alpha}\frac{\partial \hat{c}_0}{\partial E}~~,~~\frac{\partial \hat{d}_0}{\partial \kappa}=(2-\alpha)(\hat{c}_0)^{1-\alpha}\frac{\partial \hat{c}_0}{\partial \kappa},
\end{equation}
\begin{equation}
    \frac{\partial \hat{c}_0}{\partial C}=2\left(k_4\frac{\partial \hat{A}_2}{\partial C}+\hat{A}_2\frac{\partial k_4}{\partial C}\right)+k_5\frac{\partial \hat{A}_1}{\partial C}+\hat{A}_1\frac{\partial k_5}{\partial C},
\end{equation}
\begin{equation}
    \frac{\partial \hat{c}_0}{\partial E}=2\left(k_4\frac{\partial \hat{A}_2}{\partial E}+\hat{A}_2\frac{\partial k_4}{\partial E}\right)+k_5\frac{\partial \hat{A}_1}{\partial E}+\hat{A}_1\frac{\partial k_5}{\partial E}~~,~~\frac{\partial \hat{c}_0}{\partial \kappa}=2
k_4\frac{\partial \hat{A}_2}{\partial \kappa}.
\end{equation}
\end{subequations}

Next, we take derivative of~(\ref{eq:UtTransformedSeriesSolu_NNS}g) as follows
\begin{subequations}
\begin{equation}
    \frac{\partial \hat{A}_{n+3}}{\partial C} = \frac{v_3\frac{\partial u_3}{\partial C}-u_3\frac{\partial v_3}{\partial C}}{(v_3)^2}~~,~~\frac{\partial \hat{A}_{n+3}}{\partial E} = \frac{v_3\frac{\partial u_3}{\partial E}-u_3\frac{\partial v_3}{\partial E}}{(v_3)^2}~~,~~ \frac{\partial \hat{A}_{n+3}}{\partial \kappa} = \frac{\frac{\partial u_3}{\partial \kappa}}{v_3},
\end{equation}
where
\begin{multline}
    u_3 = \left\{-k_2(n+1)n-k_3(n+1)-k_1(n+1)n(n-1)\right\}\hat
    {A}_{n+1}+ \\
    \left\{-k_2(n+2)(n+1)-2k_1(n+2)(n+1)n\right\}\hat{A}_{n+2}+\hat{e}_n,
\end{multline}
\begin{equation}
    v_3=k_1(n+3)(n+2)(n+1),
\end{equation}
\begin{multline}
    \frac{\partial u_3}{\partial C}=\left\{-k_2(n+1)n-k_3(n+1)-k_1(n+1)n(n-1)\right\}\frac{\partial \hat{A}_{n+1}}{\partial C}+ \\
    \hat{A}_{n+1}\left\{-\frac{\partial k_2}{\partial C}(n+1)n-\frac{\partial k_3}{\partial C}(n+1)-\frac{\partial k_1}{\partial C}(n+1)n(n-1)\right\} + \\
    \left\{-k_2(n+2)(n+1)-2k_1(n+2)(n+1)n\right\}\frac{\partial \hat{A}_{n+2}}{\partial C} + \\
    \hat{A}_{n+2}\left\{-\frac{\partial k_2}{\partial C}(n+2)(n+1)-2\frac{\partial k_1}{\partial C}(n+2)(n+1)n\right\}+\frac{\partial \hat{e}_n}{\partial C},
\end{multline}
\begin{multline}
    \frac{\partial u_3}{\partial E}=\left\{-k_2(n+1)n-k_3(n+1)-k_1(n+1)n(n-1)\right\}\frac{\partial \hat{A}_{n+1}}{\partial E}+ \\
    \hat{A}_{n+1}\left\{-\frac{\partial k_2}{\partial E}(n+1)n-\frac{\partial k_3}{\partial E}(n+1)-\frac{\partial k_1}{\partial E}(n+1)n(n-1)\right\} + \\
    \left\{-k_2(n+2)(n+1)-2k_1(n+2)(n+1)n\right\}\frac{\partial \hat{A}_{n+2}}{\partial E} + \\
    \hat{A}_{n+2}\left\{-\frac{\partial k_2}{\partial E}(n+2)(n+1)-2\frac{\partial k_1}{\partial E}(n+2)(n+1)n\right\}+\frac{\partial \hat{e}_n}{\partial E},
\end{multline}
\begin{multline}
    \frac{\partial u_3}{\partial \kappa}=\left\{-k_2(n+1)n-k_3(n+1)-k_1(n+1)n(n-1)\right\}\frac{\partial \hat{A}_{n+1}}{\partial \kappa}+ \\
    \left\{-k_2(n+2)(n+1)-2k_1(n+2)(n+1)n\right\}\frac{\partial \hat{A}_{n+2}}{\partial\kappa} + \frac{\partial \hat{e}_n}{\partial \kappa},
\end{multline}
\begin{equation}
    \frac{\partial v_3}{\partial C}=(n+3)(n+2)(n+1)\frac{\partial k_1}{\partial C}~~,~~\frac{\partial v_3}{\partial E}=(n+3)(n+2)(n+1)\frac{\partial k_1}{\partial E}~~,~~\frac{\partial v_3}{\partial \kappa}=0,
\end{equation}
\begin{equation}
    \frac{\partial \hat{e}_n}{\partial C}=\sum_{j=0}^n \hat{A}_j\frac{\partial \hat{d}_{n-j}}{\partial C} + \hat{d}_{n-j}\frac{\partial \hat{A}_j}{\partial C},
\end{equation}
\begin{equation}
   \frac{\partial \hat{e}_n}{\partial E}=\sum_{j=0}^n \hat{A}_j\frac{\partial \hat{d}_{n-j}}{\partial E} + \hat{d}_{n-j}\frac{\partial \hat{A}_j}{\partial E},
\end{equation}
\begin{equation}
  \frac{\partial \hat{e}_n}{\partial \kappa}=\sum_{j=0}^n \hat{A}_j\frac{\partial \hat{d}_{n-j}}{\partial \kappa} + \hat{d}_{n-j}\frac{\partial \hat{A}_j}{\partial \kappa},
\end{equation}
\begin{multline}
    \frac{\partial \hat{d}_{n>0}}{\partial C}=\frac{-1}{n(\hat{c}_0)^2}\frac{\partial \hat{c}_0}{\partial C}\sum_{j=1}^n (3j-\alpha j-n)\hat{c}_j \hat{d}_{n-j} + \\
    \frac{1}{n\hat{c}_0}\sum_{j=1}^n (3j-\alpha j-n)\frac{\partial \hat{c}_j}{\partial C}\hat{d}_{n-j}+\frac{1}{n\hat{c}_0}\sum_{j=1}^n  (3j-\alpha j-n)\hat{c}_j\frac{\partial \hat{d}_{n-j}}{\partial C},
\end{multline}
\begin{multline}
    \frac{\partial \hat{d}_{n>0}}{\partial E}=\frac{-1}{n(\hat{c}_0)^2}\frac{\partial \hat{c}_0}{\partial E}\sum_{j=1}^n (3j-\alpha j-n)\hat{c}_j \hat{d}_{n-j} + \\
    \frac{1}{n\hat{c}_0}\sum_{j=1}^n (3j-\alpha j-n)\frac{\partial \hat{c}_j}{\partial E}\hat{d}_{n-j}+\frac{1}{n\hat{c}_0}\sum_{j=1}^n  (3j-\alpha j-n)\hat{c}_j\frac{\partial \hat{d}_{n-j}}{\partial E},
\end{multline}
\begin{multline}
    \frac{\partial \hat{d}_{n>0}}{\partial \kappa}=\frac{-1}{n(\hat{c}_0)^2}\frac{\partial \hat{c}_0}{\partial \kappa}\sum_{j=1}^n (3j-\alpha j-n)\hat{c}_j \hat{d}_{n-j} + \\
    \frac{1}{n\hat{c}_0}\sum_{j=1}^n (3j-\alpha j-n)\frac{\partial \hat{c}_j}{\partial \kappa}\hat{d}_{n-j}+\frac{1}{n\hat{c}_0}\sum_{j=1}^n  (3j-\alpha j-n)\hat{c}_j\frac{\partial \hat{d}_{n-j}}{\partial \kappa},
\end{multline}
\begin{multline}
    \frac{\partial \hat{c}_{n>0}}{\partial C}=(k_4n+k_5)(n+1)\frac{\partial \hat{A}_{n+1}}{\partial C}+\left(\frac{\partial k_4}{\partial C}n+\frac{\partial k_5}{\partial C}\right)(n+1)\hat{A}_{n+1}+ \\
    k_4(n+2)(n+1)\frac{\partial \hat{A}_{n+2}}{\partial C}+\frac{\partial k_4}{\partial C}(n+2)(n+1)\hat{A}_{n+2},
\end{multline}
\begin{multline}
     \frac{\partial \hat{c}_{n>0}}{\partial E}=(k_4n+k_5)(n+1)\frac{\partial \hat{A}_{n+1}}{\partial E}+\left(\frac{\partial k_4}{\partial E}n+\frac{\partial k_5}{\partial E}\right)(n+1)\hat{A}_{n+1}+ \\
    k_4(n+2)(n+1)\frac{\partial \hat{A}_{n+2}}{\partial E}+\frac{\partial k_4}{\partial E}(n+2)(n+1)\hat{A}_{n+2},
\end{multline}
\begin{multline}
    \frac{\partial \hat{c}_{n>0}}{\partial \kappa}=(k_4n+k_5)(n+1)\frac{\partial \hat{A}_{n+1}}{\partial \kappa}+\left(\frac{\partial k_4}{\partial \kappa}n+\frac{\partial k_5}{\partial \kappa}\right)(n+1)\hat{A}_{n+1}+ \\
    k_4(n+2)(n+1)\frac{\partial \hat{A}_{n+2}}{\partial \kappa}+\frac{\partial k_4}{\partial \kappa}(n+2)(n+1)\hat{A}_{n+2}.
\end{multline}
\end{subequations}

Finally, we take derivative of~(\ref{eq:A1_NNS}) as follows
\begin{subequations}
\begin{equation}
    \frac{\partial A_1}{\partial C}=\frac{1}{1-\alpha}\left(\frac{u_4}{v_4}\right)^{\frac{\alpha}{1-\alpha}}\left(\frac{v_4\frac{\partial u_4}{\partial C}-u_4\frac{\partial v_4}{\partial C}}{(v_4)^2}\right),
\end{equation}
\begin{equation}
    \frac{\partial A_1}{\partial E}=\frac{1}{1-\alpha}\left(\frac{u_4}{v_4}\right)^{\frac{\alpha}{1-\alpha}}\left(\frac{v_4\frac{\partial u_4}{\partial E}-u_4\frac{\partial v_4}{\partial E}}{(v_4)^2}\right)~~,~~\frac{\partial A_1}{\partial \kappa} = 0,
\end{equation}
where
\begin{equation}
    u_5 = k_3~~,~~v_5=A_0(k_5)^{2-\alpha},
\end{equation}
\begin{equation}
    \frac{\partial u_5}{\partial C}=\frac{\partial k_3}{\partial C}~~,~~\frac{\partial u_5}{\partial E}=\frac{\partial k_3}{\partial E}~~,~~\frac{\partial u_5}{\partial \kappa}=0,
\end{equation}
\begin{equation}
    \frac{\partial v_5}{\partial C}=A_0(2-\alpha)(k_5)^{1-\alpha}\frac{\partial k_5}{\partial C}+(k_5)^{2-\alpha}~~,~~\frac{\partial v_5}{\partial E}=A_0(2-\alpha)(k_5)^{1-\alpha}\frac{\partial k_5}{\partial E}~~,~~\frac{\partial v_5}{\partial \kappa} = 0.
\end{equation}
\end{subequations}
\nocite{*}
\bibliography{NNSPaper}
\end{document}